\documentclass[fleqn,usenatbib]{mnras}

\usepackage[T1]{fontenc}
\usepackage{ae,aecompl}
\usepackage{soul}

\usepackage{multirow}
\usepackage{float}
\restylefloat{table}

\usepackage{amsmath}	
\usepackage{amssymb}	
\usepackage{float}
\usepackage[caption = false]{subfig}
\usepackage[final]{graphicx}
 
\usepackage[normalem]{ulem}
\usepackage{scrextend}

\title[Giant collisions and formation of cold sub-Saturns]{The effect of late giant collisions on the atmospheres of protoplanets and the formation of cold sub-Saturns}

\author[Ali-Dib, Cumming, \& Lin]{
Mohamad Ali-Dib$^{1}$\thanks{E-mail: malidib@nyu.edu}, Andrew Cumming$^{2}$, and Douglas N. C. Lin$^{3,4}$ \\
$^{1}$Center for Astro, Particle and Planetary Physics (CAP3), New York University, Abu Dhabi, UAE\\
$^{2}$Department of Physics and McGill Space Institute, McGill University, 3600 rue University, Montreal, QC, H3A 2T8, Canada\\
$^{3}$Department of Astronomy and Astrophysics, University of California, Santa Cruz, CA 95064, USA\\
$^{4}$Institue for Advanced Studies, Tsinghua University, Beijing, China
}

\date{Accepted XXX. Received YYY; in original form ZZZ}

\pubyear{2021}

\begin{document}
\label{firstpage}
\pagerange{\pageref{firstpage}--\pageref{lastpage}}
\maketitle

\begin{abstract}
We investigate the origins of cold sub-Saturns (CSS), an exoplanetary population inferred from microlensing surveys. {If confirmed, these planets would rebut a theorised gap in planets' mass distribution between those of Neptune and Jupiter caused by the rapid runaway accretion of super-critical cores.} 
In an attempt to resolve this theoretical-observational disparity, we examine the outcomes of giant collisions between sub-critical protoplanets. Due to the secular interaction among protoplanets, these events may occur in rapidly depleting discs.
We show that impactors $\sim$ 5\% the mass of near-runaway envelopes around massive cores {can efficiently remove these envelopes entirely via a thermally-driven super-Eddington wind emanating from the core itself, in contrast with the stellar Parker winds usually considered.} 
After a brief cooling phase, the merged cores resume accretion.  But, the evolution timescale of transitional discs is too brief for the cores to acquire sufficiently massive envelopes to undergo runaway accretion despite their large combined masses. Consequently, these events lead to the emergence of CSS without their transformation into gas giants.  
We show that these results are robust for a wide range of disc densities, grain opacities and silicate abundance in the envelope. Our fiducial case reproduces CSS with heavy ($\gtrsim 30 M_{\oplus}$) cores and less massive (a few $M_\oplus$) sub-critical envelopes. 
We also investigate the other limiting cases, where continuous mergers of comparable-mass cores yield CSS with wider ranges of core-to-envelope mass ratios and envelope opacities. Our results indicate that it is possible for CSS and Uranus and Neptune to emerge within the framework of well studied processes and they may be more common than previously postulated.
\end{abstract}




\begin{figure*}
\begin{centering}
        \includegraphics[scale=0.50]{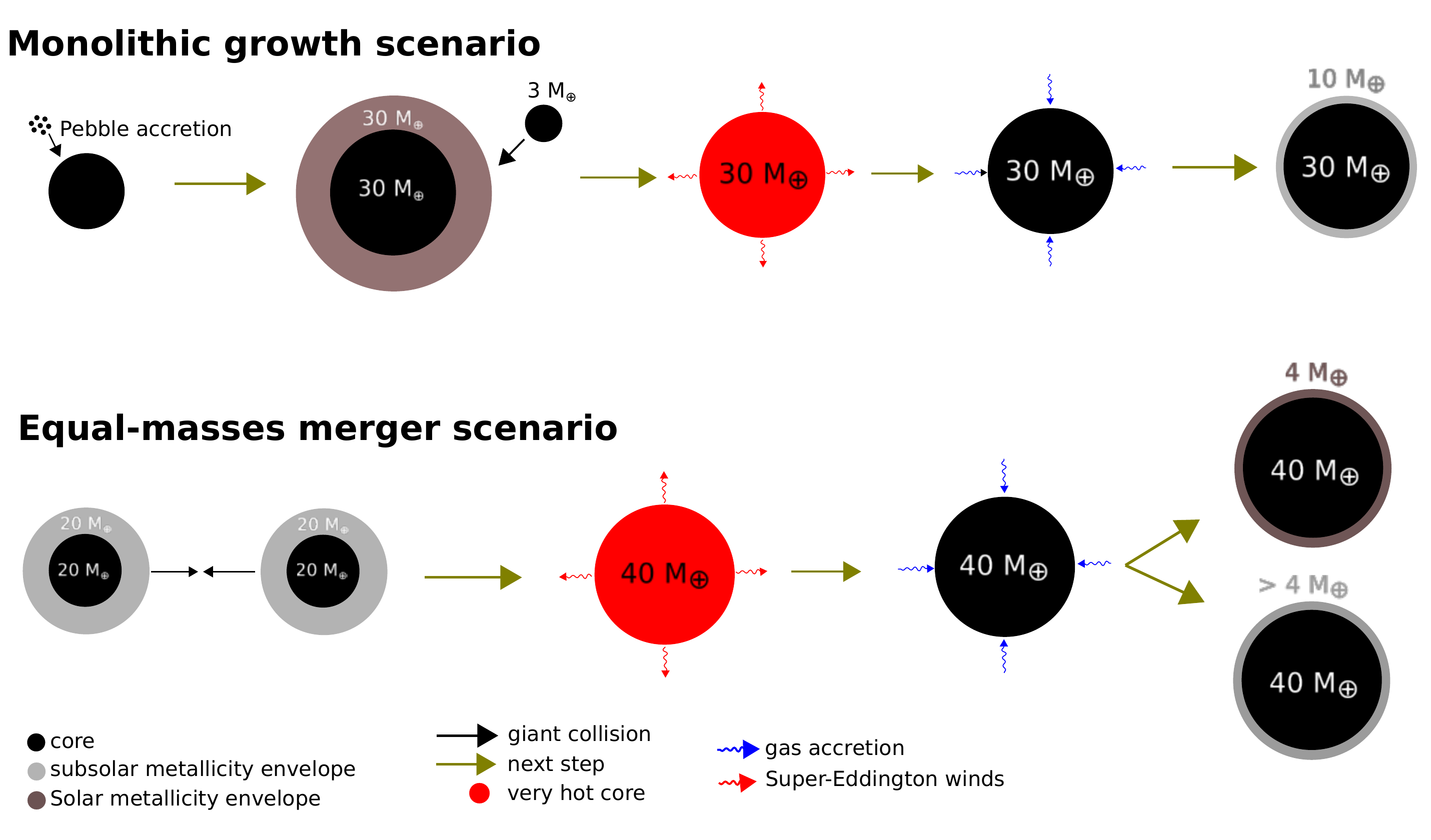}
   \caption{A schematic showing possible cold sub-Saturn (CSS) planet formation channels. 
   The first scenario (top) represents the case of a low mass impactor colliding with a $\sim 30 M_{\oplus}$ core that was formed through pebble accretion in part of the disc with large isolation mass and accreted a near-critical envelope before the collision. The collision strips the envelope entirely, and subsequently the core cools down and re-accretes a thin, subcritical envelope. The second scenario (bottom) represents the case of a collision between equal-mass planets, where for example two $\sim 15$--$20 M_{\oplus}$ cores with near-critical envelopes merge, stripping their envelopes in the process. The product of the collision is a $\sim 30$--$40 M_{\oplus}$ core that then proceeds to cool and reaccrete an envelope.
   In the first scenario, high pre-merger core masses necessitate solar envelope metallicity to avoid runaway accretion before the collision. This is not the case for the lower mass cores in the second scenario. }
    \label{fig:cartoon}
    \end{centering}
\end{figure*}

\section{Introduction}
\label{sec:introduction}
Recent exoplanet surveys have revealed an astonishing diversity in planetary systems that challenges formation models \citep{demographics}. While the highly successful transit method is sensitive to planets on short periods \citep{transit}, microlensing surveys probe planets on significantly wider orbits \citep{microlensing, Guerrero2021}. One of the most striking results from microlensing surveys is the statistical inference of a large population of cold sub-Saturn-mass planets at orbital radii of a few ${\rm AU}$ (we refer to these as CSS for the rest of this manuscript). {Although based on only a few detections, analysis of the survey sensitivities suggests that these planets with host-star mass-ratio $q$ between $10^{-4}$ and $3 \times 10^{-4}$ ($\approx 30$--$90\ M_\oplus$ for a solar mass star), are common \citep{suzuki2016, Jung2019}. Using a limited number of detections, \cite{suzuki2016} constrained the bias-corrected occurrence rate of CSS to $\sim$ 0.1-0.5 per star. They moreover found a ``break'' in the slope of the distribution around $\log(q) \sim -3.7$, indicating a paucity of lower mass planets. \cite{Jung2019} on the other hand, with one new CSS detection, found the break to be significantly steeper, and around $\log(q) \sim -4.25$. They moreover concluded that accounting for observational biases had little effects on these results. Finally, using an automated ``Anomaly Finder'' algorithm applied to archival data, \cite{hwang2021} found no slope change in the distribution at all.}
{However, the observed mass-distribution statistics may be significantly modified, in favor of planets below this mass range, by the 
ongoing systematic searches \citep{2021arXiv210311880Z} which reveal some tantalizing signatures of a much larger population of previously undetected Neptune-mass planets \citep{Hwang2021}.}

If confirmed by future observations, including direct imaging, radial velocity, and WFIRST surveys, CSS would defy classic formation models predicting that their progenitors should have grown into Jupiter-mass planets \citep{microl1,microl2,microl3}. \cite{microl1} in particular showed that detailed population synthesis models \citep{ida, morda, ida2013} predict an order of magnitude fewer CSS planets than observed with the  early microlensing surveys. 
This theoretically-predicted ``desert'' of intermediate-mass long-period planets \citep{ida} is mainly caused by the cooling-accretion timescales of envelopes around cores being 1) inversely proportional to their total mass and 2) {shorter than the 1-10 Myr (with a median of 3 Myr) persistence timescale of protostellar discs \citep{disklife1,disklife2,disklife3}.} The envelopes hence can rapidly contract, quickly grow to become comparable to the cores in mass, undergo runaway accretion, and transform into gas giants in the process. This tendency is especially acute for CSS at intermediate semimajor axis where the disc gas density is still relatively large and its temperature is low. Eventually, as the mass of the gas giant planet grows, the tidal interaction with the disc induces gap formation \citep{goldreichtremaine1980,linpap}, which can can possibly quench the gas supply and limit the planet's mass \citep{linpap1980, linpap93, bryden1999,ida, lichenlin2021}, although uncertainties on whether this is possible remain \citep{morby}.

{ Gaps have been commonly found in the ALMA images of protoplanetary 
discs\citep[e.g.][]{andrews2018,Huang2018,long2018,long2019,lodato2019}. Their observed width have been 
applied to infer the mass of the planets and derive a continuous mass distribution for 
a rich population of hypothetical planets in the range of several Earth to Jupiter masses
\citep{lodato2019,Nayakshin2019}. However, the gap separation of Neptune-mass planets with small 
eccentricities is comparable to that induced by Saturn-mass planets on circular orbits. This degeneracy
may obliterate the discrepancy between the theoretical bi-modal mass distribution and the observed
continuous gap width distribution \citep{chen2021}.}

The discrepancy between the predicted dip in the mass distribution and the observational inference of a continuous mass function can also be resolved if either the 
disc is timely and rapidly depleted during the emergence of CSS and gas giants, or if the envelope cooling and accretion rate are suppressed during some critical growth stage.  
In an attempt to explore these potential scenarios for the origin of CSS in the context of the
core-accretion planet-formation model, we investigate here the consequence of giant collisions taking place during the
dynamically-active dispersal phase of the disc. We address three physical effects induced by giant collisions: 1) the vast amount of energy
dissipation, which contributes to the loss of the protoplanet's gaseous envelope, and growth in core mass through mergers, 
2) the heavy-element contamination and opacity enhancement of the immediate surroundings which lowers the cooling rate, promotes a vapor-to-condensate phase transition, and reduces the quasi-hydrostatic contraction rate in the re-acquired envelope, and 3) the rapid depletion of the ambient gas in the natal disc.  

Using a simplified post-collision envelope evolution model, we start in \S\ref{sec:giantimpact} by showing that giant collisions can  efficiently ``reset'' a sub-critical envelope to a pre-accretion state by thermally stripping it almost completely. 
With their reduced post-collision total masses, these residual cores need to replenish a substantial amount of gas in their envelope to reach but not exceed the CSS's mass. In \S\ref{sec:envelopeaccretion} we present a grid of 1D envelope cooling-accretion models for different ranges of a) disc density, b) disc depletion time scale, c) with or without entropy advection to and from the cores' proximity, d) opacity, and e) dry or wet adiabat for the heavy-element-enriched cooler outer and middle zones in the planetary envelope. This investigation provides some constraints on the magnitudes of physical parameters that are suitable for the formation of CSS through giant collisions during advanced stages of disc evolution.  

We interpret these results and propose a plausible formation scenario for intermediate-mass planets in \S\ref{sec:icegiants}. The basic picture is that giant collisions strip massive cores of their envelopes as the protoplanetary disc depletes, and under the right conditions, they can re-accrete a low mass envelope that leaves the planet with a mass in the sub-Saturn range. Fig. \ref{fig:cartoon} provides a schematic illustration for two limiting cases of a large core with a small impactor or two equal mass cores colliding. In \S \ref{sec:icegiants}, we also discuss the formation of ice giants, including Uranus and Neptune, which contain envelopes with masses comparable to that of the Earth and ten-times more massive cores.  We summarize our results and discuss their implications in \S\ref{sec:summary}.

\section{Effect of giant collisions on the protoplanetary envelope}
\label{sec:giantimpact}

Giant collisions are thought to play an integral role in the formation and evolution of the solar system and exoplanetary systems. 
They have long been invoked as a mechanism to grow planetary cores, strip their envelopes, { alter their spin,} and affect their dynamics \citep{linida1997, lietal2010, liu2015c, Ogihara}. In the present context, these collisions lead to two additional consequences: 1) they can reset the protoplanetary envelope to an entropy
state corresponding to a much earlier epoch (see below), and 2) they can significantly increase the metallicity of the envelope.
These collisions 
are most likely to occur during the advanced stages of disc evolution, particularly during the transitional disc phase, when the gas 
density is low and the disc dispersal timescale is reduced to $10^5$ yr \citep{zhou2007, ida2013, ginzburg2020}. 

In this section, we put forward a simplified post-collision evolution model for the protoplanet and its envelope.
{ First, we discuss the probability of having these giant collisions at a few AU.  In conventional core-accretion 
scenarios, planetesimals undergo oligarchic growth to become protoplanetary embryos\citep{kokuboida1998, ida}.
Mutual interaction between these embryos excites their eccentricity \citep{chambers}. 
When the embryos' velocity dispersion become comparable to their surface escape speed, their collisional cross section
is reduced to their physical size \citep{safranov}.  For a minimum mass nebular model, Earth-mass embryos collide 
and merge with each other on a timescale $\tau_{\rm collide}$ several dozen Myr in the terrestrial-planet 
domain \citep{chamberswetherill}. These results match well with N-body simulations on the final growth of modest-mass embryos \citep{Kokubo2006}. The magnitude of $\tau_{\rm collide}$ is even larger beyond the snow 
line \citep{goldreichetal2004}, posing a potential dichotomy between the formation probability of gas planets in a minimum
mass solar nebula before its depletion \citep{pollack} and their observed occurrence rate around solar type stars \citep{kunimoto}. 

However, these early estimates and simulations are mostly relevant for local interaction between planetesimals and 
embryos in a minimum mass solar nebula. There are several aspects which may markedly reduce both orbit crossing 
($\tau_{\rm cross}$) and collision ($\tau_{\rm collide}$) time scales.
Through analytic calculation and extensive N-body simulations, \cite{zhou2007} determine 
$\tau_{\rm cross}$ for systems containing multiple equal-mass embryos. They show $\tau_{\rm cross}$ is a rapidly 
increasing function of the normalized spacing (in terms of their Hill's radius) between the embryos and it decreases 
with their eccentricity.  For example, in densely-packed protoplanetary discs containing several 15 M$_{\oplus}$ 
embryos separated by 1 AU starting at 5 AU, with eccentricities of 0.3, we estimate a $\tau_{\rm cross}$ is 
comparable to the lifetime of the transitional disc $\sim 10^5$ yr. [Similar effects also lead to dynamical 
relaxation and migration of gas giant planets \cite{rasio, weidenschilling, linida1997, zhou2007, Juric}.]
Moreover, $\tau_{\rm collide} \propto \Sigma_{\rm embryo}^{-1}$, embryos' mass 
surface density \citep{nagasawa2005} and the magnitude of $\Sigma_{\rm embro}$ can be much larger than that of 
the minimum mass solar nebula either globally in building-block-rich discs or near migration traps where embryos 
converge \citep{liuetal2015}.  The cross section for cohesive collisions is also enlarged from that of bare cores
for those surrounded by contracting envelopes.  Prescriptions of these effects have been applied to population 
synthesis models to reproduce the observed mass and spacial distribution of exoplanets, including gas giant planets \citep{idalin2010, 
ida2013}.  During the final stage of their formation, gas giant planets' rapidly increasing gravitational 
perturbation de-stabilizes the orbits of nearby embryos and enhances their collision frequency and probability
\citep{zhoulin2007}.  In transition discs, the propagation of gas giants' secular resonances (due to their distant
perturbation rather than close encounters) leads to the alignment
of residual embryos' longitudes of periapse, reduction of their relative speed, and increase in their collision 
frequency by an order of magnitude 
\citep{nagasawa2005, thommes2008, sec1,sec2}. Based on these considerations, we adopt the assumption that giant 
impacts between protoplanetary embryos are not only an effective pathway for the emergence for gas giants but 
also synonymous with their presence.}

\begin{figure*}
\begin{centering}
        \includegraphics[scale=0.40]{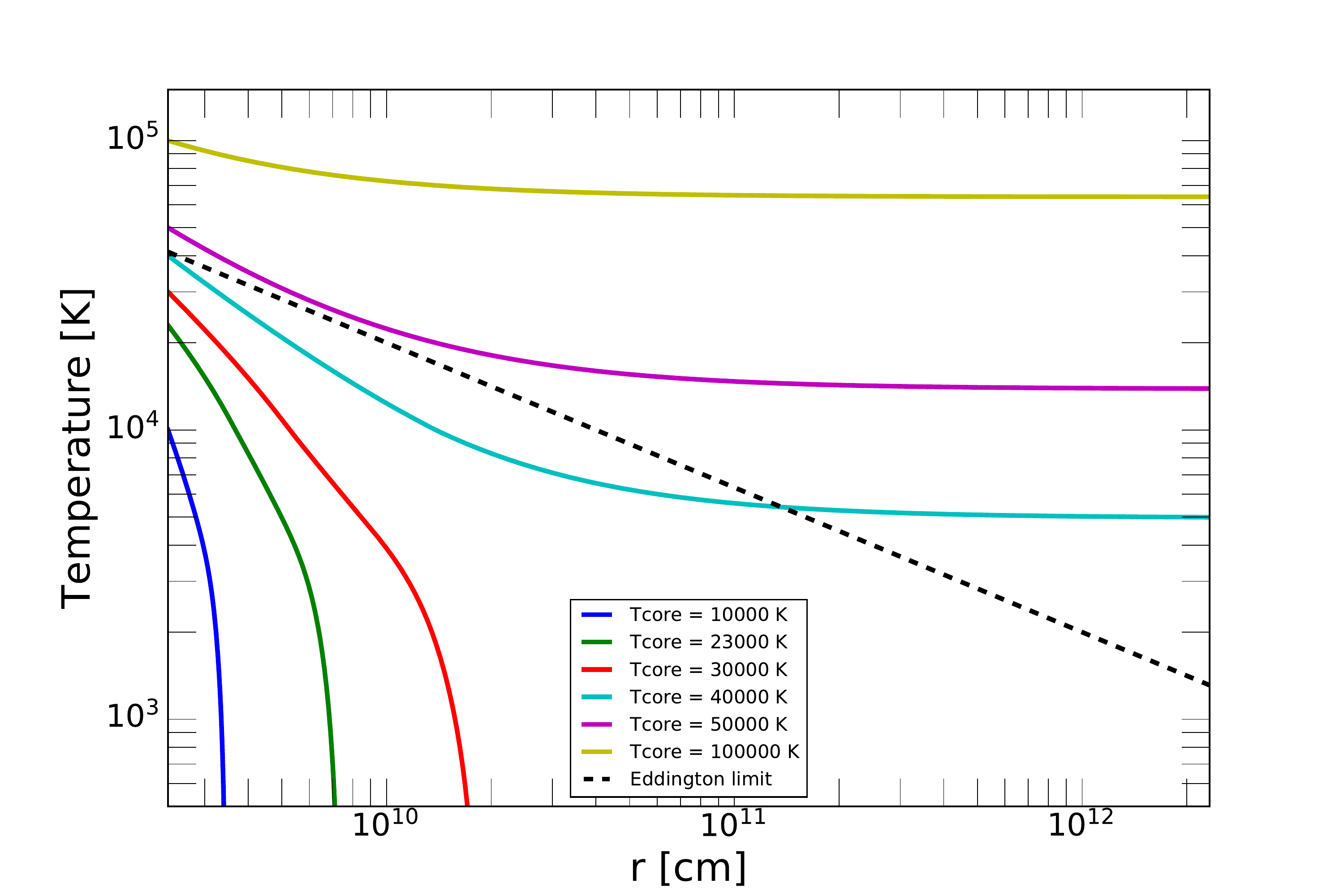}
   \caption{The temperature profiles of an envelope dominated by the very hot adiabat of the core post-collision, as a function of the core's temperature. The dashed black line is the Eddington temperature. For very hot cores, the envelope is shed thermally through Eddington winds. {Reaccreting an envelope necessitates} waiting for the core to cool down below a threshold value. }
    \label{fig:planets}
    \end{centering}
\end{figure*}

\subsection{Loss of the envelope during giant impacts and post-collision core temperature}
\label{sec:envloss}
Multi-component simulations show that giant impacts lead to a substantial atmospheric mass loss, intense dissipation of kinetic energy into heat, and the expansion of the residual envelope \citep{genda2003, liu2015b}. After the impact, the envelope loses a substantial amount of gas and then readjusts to a new series of quasi-hydrostatic equilibria as it cools, contracts, and accretion of disc gas resumes.

As the collisional cross section is a function of the Hill radius $R_{\mathrm{H}}=a\left(M_{\mathrm{tot}} / 3 M_{\star}\right)^{1 / 3}$, 
the collision probability is highest for massive protoplanets at large semimajor axis. Therefore, for our fiducial model we consider a planet just before reaching runaway accretion, with approximately equal envelope and core masses $M_{e} \sim M_{c} \sim 30 M_{\oplus}$, 
and a total mass $M_\mathrm{tot} \sim 60$ M$_{\oplus}$. This value of $M_{c}$ is a conservative choice of core mass for estimating mass loss, since lower values would lead to 
a hotter post-collision core and thus more substantial envelope mass loss.  

\cite{collisions} investigated the amount of atmospheric mass loss following a giant collision resulting from the shock wave launched in the atmosphere by global ground motion. They found that the global mass loss fraction for an adiabatic atmosphere $X_{\text {loss}}$ is given by
\begin{eqnarray}
\label{eq:masslossfrac}
X_{\text {loss}}&=&0.4\left(\frac{v_\mathrm{imp} M_\mathrm{imp}}{v_\mathrm{esc} M_\mathrm{tot}}\right)+1.8\left(\frac{v_\mathrm{imp} M_\mathrm{imp}}{v_\mathrm{esc} M_\mathrm{tot}}\right)^{2}\nonumber\\&&-1.2\left(\frac{v_\mathrm{imp} M_\mathrm{imp}}{v_\mathrm{esc} M_\mathrm{tot}}\right)^{3},
\end{eqnarray}
where $M_\mathrm{imp}$ and $v_\mathrm{imp}$ are the impactor's mass and velocity,  $M_\mathrm{tot}$ is the planet's total (core + envelope) mass, and $v_\mathrm{esc}$ the planet's escape velocity. Similar descriptions exist for isothermal atmospheres. Assuming $M_\mathrm{imp}=3 M_\oplus$, and $v_\mathrm{imp}$/$v_\mathrm{esc}$ = 1, equation (\ref{eq:masslossfrac}) gives $X_{\text {loss}} =$ 0.06. Therefore, even for a very energetic collision, the mass loss fraction due to the shock wave is still modest. {These values are consistent with the results of \cite{inamdar} when considering the same envelope mass and collision velocity.}

{However, since the simulations of \cite{collisions} and \cite{inamdar} are hydrodynamic in nature, they focus on the shock propagation physics and do not account for the thermal interactions between the core and the remaining envelope during the post-collision evolution. We hence present a simplified prescription to account for these effects. We emphasize that our subsequent analysis is fundamentally different than that presented by \cite{liu2015b}, who accounted for the post-collisional thermal interaction of the envelope with the stellar-driven Parker winds. In this paper, on the other hand, while we ignore these effects as they are negligible at 5 AU (the analysis of \cite{liu2015b} was for 0.1 AU), we include the thermal irradiation effects of the core itself that was heated up significantly by the collision. }




Assuming that the collision is inelastic, and that the remaining energy is deposited entirely in the solid core \citep{liu2015b}, we can estimate the core's post-collision energetic state by {subtracting the gravitational binding energy of the lost fraction of the envelope ($E_g^L$) from the kinetic energy of the impactor ($K_\mathrm{imp}$). }
{Assuming $K_\mathrm{imp} >> E_g^L$ (as shown in sections \ref{mmim} and \ref{mmim2} to be the case for our parameters), we then estimate the core's post-collision temperature to be:} 
\begin{equation}
\label{eq:tcedd}
T_c = T_{c,0} + \frac{(K_\mathrm{imp} - E_g^L)}{(M_{c} + M_\mathrm{imp}) C_V} \sim T_{c,0} + \frac{K_\mathrm{imp}}{(M_{c} + M_\mathrm{imp}) C_V}
\end{equation}
where $T_{c,0}$ is the core's pre-collision temperature. This is the temperature at the boundary between the core and the base of the planetary envelope. We take the heat capacity to be $C_V = 1129 \ \mathrm{J} \  \mathrm{kg}^{-1} \mathrm{K}^{-1}=1.129\times 10^7\ {\rm erg\ g^{-1}\ K^{-1}}$ from \cite{boujibar} for silicates. We assume a core density $\rho_c =$ 3.3 $\mathrm{g\ cm^{-3}}$. 

With the same parameters for the impact used above ($M_\mathrm{imp}=3 M_\oplus$, and $v_\mathrm{imp}$/$v_\mathrm{esc}$ = 1) and taking $T_{c,0} = 2\times 10^4$ K (equal to the pre-collision envelope base temperature), we find the post-collision core temperature to be T$_c\sim 3.5\times10^5$ K. The core's post-collision luminosity is therefore estimated to be $L_c \sim 4\pi R_c^2 \sigma T_c^4\sim 3\times10^{37}$ erg s$^{-1}$, where the core's radius $R_c$ is calculated from its density and mass. {A core density 3 times the value assumed here would result in $L_c$ only a factor 2 smaller. }

\subsection{Post-impact thermal structure of the envelope}

We analyze the post-collision evolution under the assumption that the merger products fully retain their cores and lose most of their pre-collision gas content.  On a short dynamical time scale, a low-mass envelope is retained with a re-established hydrostatic and quasi-equilibrium thermal structure. Subsequently, the density and temperature distributions in the envelope evolve as entropy is lost near its outer boundary on a longer heat transfer timescale. 

For the high core temperature $\sim 10^5\ {\rm K}$ and luminosity that we found in \S\ref{sec:envloss}, the envelope structure is fundamentally different from that of a classical embedded protoplanet. In this case, the envelope becomes fully convective after a thermal quasi-equilibrium is established such that the luminosity throughout the envelope is approximately equal to the core's emergent luminosity. Since the core's temperature and energy content are significantly higher than those of the local disc, the entire envelope attains the same adiabat as that of the core.  This thermal structure is in contrast to the classical adiabatic envelopes around their cooler progenitors which matches onto the adiabat of the cold disc gas. The density profile, and thus mass, of an adiabatic envelope are governed entirely by the boundary condition and adiabatic index.   After the collision, a large elevation in entropy reduces the density and mass throughout the envelope.  Consequently, the envelope surrounding the core expands beyond the Hill radius and sheds any mass in excess of that allowed by the adiabatic solution. The outcome of the collision is therefore a very hot, adiabatic envelope with a much reduced mass. Subsequently, the core and the envelope have to lose entropy and cool down enough to allow the re-establishment of an outer radiative zone before the merger product can resume its accretion of the disc gas.

\begin{figure*}
\begin{centering}
        \includegraphics[scale=0.40]{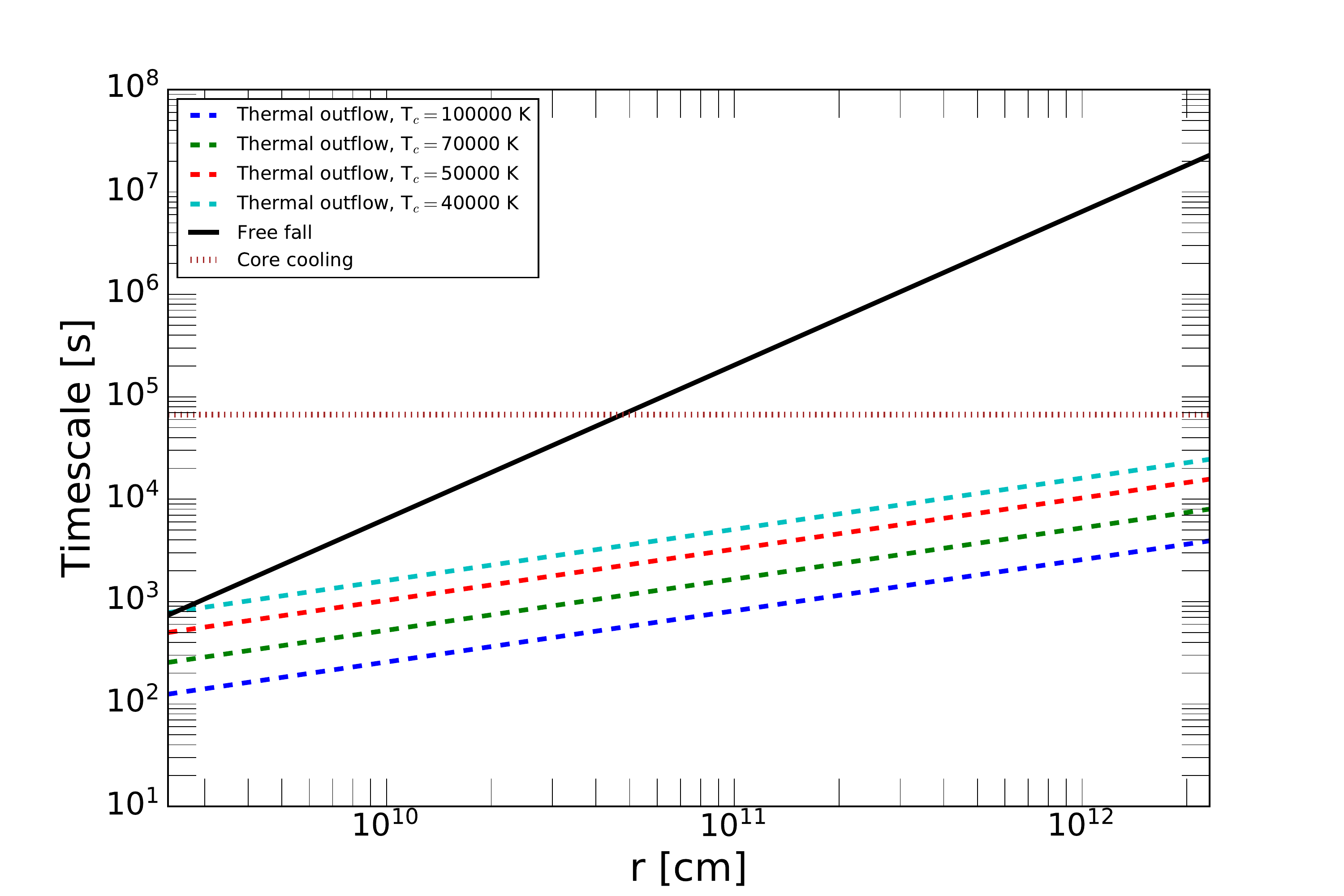}
   \caption{A comparison of the free fall timescale (${t}_{d}=\left(r^{3} / G M_{\text {core }}\right)^{1 / 2}$), the core's  initial cooling timescale, and the radiation-pressure-driven expansion timescale ($t_{\rm exp} = r / v_{\rm exp}$) for different core temperatures. All of the parameters are the same as in Fig. \ref{fig:planets}. }
    \label{fig:tdyn}
    \end{centering}
\end{figure*}

The thermal structure of an adiabatic envelope in hydrostatic balance is described by
\begin{equation}
\frac{\mathrm{d} P}{\mathrm{d} r} =-\frac{G M_{c}}{r^{2}} \rho(r) \ \ \ \ \ \ {\rm and} \ \ \ \ \ \
\frac{\mathrm{d} T}{\mathrm{d} r} =\nabla_\mathrm{ad} \frac{T}{P} \frac{\mathrm{d} P}{\mathrm{d} r},
\end{equation}
where $\nabla_\mathrm{ad}$ is the adiabatic gradient, and since we expect the mass of the hot envelope to be much smaller than the core mass, we neglect the envelope's contribution to the gravity by using $M_{c}$ instead of $m(r)$. 
Taking an ideal gas equation of state with mean molecular weight $\mu$, we then find
\begin{equation}
{dT \over dr} = -\nabla_\mathrm{ad} \frac{GM_{c}}{r^2}\frac{\mu}{k_B},
\end{equation}
which has the analytical solution
\begin{equation}\label{eq:Tanal}
T(r) 
= T_c - \nabla_\mathrm{ad} \frac{G M_{c} \mu}{k_BR_c}\left[1-{R_c\over r}\right].
\end{equation}
This solution is plotted for different core temperatures in Fig.~\ref{fig:planets}.

The temperature profiles in Fig.~\ref{fig:planets} show two different behaviors. For
\begin{eqnarray}\label{eq:threshold}
T_c>\nabla_\mathrm{ad} \frac{G M_{c} \mu}{k_B R_c}  &=& 2.3\times 10^4\ {\rm K}\ \left({M_{c}\over 30M_\oplus}\right)\\&&\times\left({R_c\over 3R_\oplus}\right)^{-1}\left({\nabla_\mathrm{ad}\over 0.3}\right)\times {\mu}
\end{eqnarray}
the temperature remains much larger than the disc temperature at large radius. This high-temperature phase represents a transient state, as these envelopes are dynamically unstable and they cool down by shedding the hot gas back to the disc via super-Eddington winds (as discussed in the next section).

Once the core has cooled below the threshold value in equation (\ref{eq:threshold}), the envelope temperature in our adiabatic solutions decreases sharply with radius. The atmosphere in this case has a finite thickness (eq.~\ref{eq:Tanal} shows that $T$ goes to zero at a finite radius). This result is a consequence of our assumption that the temperature gradient is adiabatic. Actually, in this low-temperature limit, a radiative zone would develop in the outer envelope and regulate the thermal interactions between the envelope and the disc.





\subsection{Super-Eddington winds}
\label{sew}
The hot envelope structure described above with a high luminosity core and significant radiation pressure can lead to super-Eddington winds. The Eddington luminosity is
\begin{equation}
L_{\mathrm{Edd}}=\frac{4 \pi G M_{c} m_{\mathrm{p}} c}{\sigma_{\mathrm{T}}} \approx 10^{34}\ {\rm erg\ s^{-1}}\ \left({M_{c}\over 30M_\oplus}\right),  
\end{equation}
where $\sigma_{\mathrm{T}}$ is the Thompson scattering cross section (appropriate for temperatures above the grain condensation temperature).
The temperature at which the core luminosity $L_c$ exceeds the Eddington limit is
\begin{eqnarray}
\label{teddeq}
T_\mathrm{Edd} &=& \left( {GM_{c} m_{p} c \over \sigma r^2 \sigma_T} \right)^{1/4}\\&\approx& 2.5\times 10^4\ {\rm K} \ \left({M_{c}\over 30M_\oplus}\right)^{1/4}\left({r\over 10R_\oplus}\right)^{-1/2}.\nonumber
\end{eqnarray}

This temperature can then be compared to the thermal profile calculated above, as shown in Fig. \ref{fig:planets}. In the early phases, 
when $T(r)\gg T_\mathrm{Edd}$, the outer envelope is hot enough for the radiation pressure to overcome gravity, pushing the hot gas outside 
of the Hill radius, and possibly opening a gap just inside of it.  The envelope gas is accelerated to a
radiation-pressure-driven expansion speed $v_{\rm exp}$ calculated, when neglecting gas pressure, as 
\begin{equation}
v_{\rm exp} \frac{dv_{\rm exp}}{dr} \simeq \frac{L}{4\pi r^2}\frac{\kappa}{c} \ \ \ \ \Rightarrow \ \ \ \ v_{\rm exp} \simeq \left( {L \kappa \over 
2 \pi c R_c} \right)^{1/2}
\end{equation}
where in this case $\kappa$ is the Thompson scattering opacity, and the envelope luminosity $L = 4\pi \sigma r^2 T_{c}^4$.
We solve this equation numerically in Fig.~\ref{fig:tdyn}, and show that the expected velocities are considerably larger than the residual 
core's escape speed $v_{\rm esc} (M_{c}) = (G M_{c}/R_c)^{1/2}$. 

Fig.~\ref{fig:tdyn} shows moreover that the inertia of the escaping envelope enables it to reach the residual core's Hill's radius (with an expansion factor $R_H (M_{c})/R_c = (a/R_\ast) (\rho_c/3 \rho_\ast)^{1/3} \sim (a/R_\ast) \sim 10^3$ 
where $\rho_\ast$ is the star's internal density) on a time scale $t_{\rm exp} (R_H) = R_H/v_{\rm exp} \sim 10^4$s for $T_c$ in the range of 
$4 \times 10^4$--$10^5$K.  This radiation-pressure-driven outflow time scale is much shorter than both the dynamical free-fall time scale at the Hill radius $t_H = (R_H^3/G M_{c}) ^{1/2} \simeq t_{\rm cool} (a/R_\ast)^{3/2} \sim 7 \times 10^7$ s, and the planet's cooling 
timescale (for both its core and envelope) which initially is $t_{\rm cool}\sim  E_{c} / L \sim 2 \times 10^{42}\ {\rm erg}/3 \times 10^{37}\ {\rm erg \ s^{-1}} \sim 6.7 \times 10^4$ s where $E_c$ is the thermal energy of the core.
In Appendix \ref{sec:marginaleddington}, we discuss a short-lived intermediary ``marginal Eddington'' phase during which the core's radiation pressure shrinks its Hill radius, resulting in even shorter outflow timescale than estimated above.
This comparison implies that the planet's pre-collision envelope is ejected before its post-collision envelope is significantly cooled with declining temperatures. When the core's temperature decreases below the threshold value discussed in the previous section, the radiation pressure throughout the entire envelope falls below the Eddington limit. This transitional state corresponds to a luminosity $\sim 10^{33}$ erg/s for which a radiative zone develops in the envelope.


\begin{table*}
\begin{tabular}{|l|c|c|c|c|c|c|c|}
\hline
Model parameters                         & $E_G$ [erg]       & $U$ [erg]     & |$E_G + U$|         & $E_G^{a1}$       & $E_G^{a2}$   &  $m_\mathrm{imp}^{\mathrm{min},3}$ [$M_{\oplus}$] &  $m_\mathrm{imp}^\mathrm{min}$ [$M_{\oplus}$] \\ \hline
$M_c$=30 $M_\oplus$, $\gamma$=1.40 & -3.04 $\times 10^{41}$ & 2.12 $\times 10^{41}$ & 9.14 $\times 10^{40}$ & -7.74 $\times 10^{40}$ & -8.3 $\times 10^{41}$ & 1.50 & 1.50  \\ \hline
$M_c$=30 $M_\oplus$, $\gamma$=1.25    & -4.94 $\times 10^{41}$ & 4.72 $\times 10^{41}$ & 2.22 $\times 10^{40}$ & -5.76 $\times 10^{40}$ & -8.3 $\times 10^{41}$ & 0.43 & 0.55 \\ \hline
$M_c$=20 $M_\oplus$, $\gamma$=1.40 & -1.71 $\times 10^{41}$ & 1.18 $\times 10^{41}$ & 5.31 $\times 10^{40}$ & -5.01 $\times 10^{40}$ & -4.22 $\times 10^{41}$  & 1.15 & 1.15 \\ \hline
$M_c$=20 $M_\oplus$, $\gamma$=1.25 & -2.68 $\times 10^{41}$ & 2.51 $\times 10^{41}$ & 1.78 $\times 10^{40}$ & -4.10 $\times 10^{40}$ & -4.22 $\times 10^{41}$  & 0.38 & 0.53 \\ \hline
\end{tabular}

\caption{Contributions to the binding energy of the pre-collision envelope compared with the kinetic energy of the impactor. Column 1: model parameters. In all cases the values are reported for $M_e=M_c$.  Columns 2, 3, and 4 : the envelope's gravitational, internal, and total energies calculated numerically. Columns 5 and 6: The gravitational binding energy of the envelope calculated analytically with $E_G^{a1}$ and $E_G^{a2}$ assuming the mass to concentrate near the {radiative-convective boundary (RCB)} and the core, respectively. Column 7: the impactor's  minimal mass (with v$_\mathrm{imp}$ = v$_\mathrm{esc}$) needed for its kinetic energy to equal the envelope's total energy (column 4). Column 8: the impactor's minimal mass when accounting for conditions 1 and 2. }
\label{tablenummimm}
\end{table*}

\subsection{The minimum impactor mass}
\label{mmim}
\subsubsection{Condition 1: hydrodynamic constraints}
There must be a minimum impactor mass required to thermally reset the envelope. One consideration for our calculations is that for equation~(\ref{eq:masslossfrac}) for the envelope mass loss fraction from \cite{collisions} to be applicable, the radius of the impactor needs to be larger than the scale height of the envelope at the core interface, such that its mass must be larger than the air-mass it encountered before reaching the core.  For a realisable pre-impact configuration -- i.e. an embedded progenitor in a typical protostellar disc with a marginal mass and structure to avoid runaway gas accretion over $\sim 1$--$2$ Myr, we consider a model for a $M_{c}=$30 M$_{\oplus}$ core with a slightly sub-critical envelope {\color {black} (with a mass $M_{e} \sim 30
M_\oplus$)}.  In this case, the scale height at the core level is $H \sim 6.2\times10^8$ cm. For a 
rock-ice core with $\rho = 3.3$ g/cm$^3$, an impactor {\color {black} with a radius $\sim H$} has a mass of 0.55 M$_{\oplus}$. 

{A closely related consideration is the} effect of aerodynamic drag on the impactor. We can define the drag force in the large Reynolds number limit as :
\begin{equation}
    F_D = 0.44 \pi R_\mathrm{imp}^{2} \rho_g \frac{v_\mathrm{imp}^{2}}{2}
\end{equation}
Assuming the same 0.55 M$_{\oplus}$ impactor scenario with v$_\mathrm{imp}$ = v$_\mathrm{esc}$, and for $\rho_g = 0.1\ {\rm g\ cm^{-3}}$ at the core level, we find that $F_D / F_g \ \sim $0.034, where $F_g$ is the protoplanet-impactor gravitational force. We can hence safely ignore drag effects for an impactor that satisfies condition 1.  

\subsubsection{Condition 2: thermal constraints}
The second condition needed to reset the envelope is the onset of the super-Eddington winds discussed above. For this mechanism to operate, the impactor's remaining energy dissipated in the core needs to be sufficiently large to raise its temperature to $T_{\rm Edd}$ (eq. \ref{teddeq}). 
From equation (\ref{eq:tcedd}) above, and assuming again v$_\mathrm{imp}$ = v$_\mathrm{esc}$,  we can write
\begin{equation}
M_\mathrm{imp} \sim \frac{(T_{\rm Edd} - T_{c,0}) \times C_V R_c}{G}\frac{M_c}{(M_c+M_e)} 
\end{equation}
This is $\sim$ 0.5 $M_{\oplus}$ for $M_{c}=M_{e}=$30 M$_{\oplus}$, which is comparable to condition 1. 

This condition can be extended further by requiring that, as discussed in section \ref{sew}, the thermal outflow timescale at the Hill radius $t_\mathrm{exp} (R_H)$ be smaller than the core cooling timescale $t_\mathrm{cool}$. This is summarized as $t_\mathrm{exp} (R_H) \sim t_\mathrm{cool} \sim E_c/L \sim K_\mathrm{imp}/L$.  We hence write
\begin{equation}
 R_H \sqrt{\frac{2\pi c R_c}{L \kappa}} = \frac{K_\mathrm{imp}}{L},   
\end{equation}
and adding v$_\mathrm{imp}$ = v$_\mathrm{esc}$, we finally get
\begin{equation}
    M_\mathrm{imp} = \frac{R_H R_c}{(M_c+M_e)G}\sqrt{ \frac{2\pi c L R_c}{\kappa_\mathrm{Edd}}}.
\end{equation}
For $T_c = 4\times 10^4$ K (for which the thermal outflow timescale is longest), and 30 M$_{\oplus}$, we get $M_\mathrm{imp} = $ 0.12 M$_{\oplus}$.

\subsubsection{Condition 3: energetic constraints}
\label{mmim2}

The usually most stringent condition to reset the envelope through a giant collision is that the impactor's energy deposited into the core needs to be larger than the binding energy of the entire remaining envelope. More generally, this condition can be stated by requiring that the kinetic energy of the impactor be larger than the total energy of the envelope pre-collision. If this condition is not met, then the core's luminosity will simply decrease below the Eddington limit after shedding a small fraction of the remaining envelope. 

We hence require that $K_\mathrm{imp} > E_G + U$ where $E_G$ and $U$ are the envelope's gravitational and internal energies defined in equations (\ref{egeq}) and (\ref{ueq}). We start by calculating these terms numerically using the model presented in \S \ref{sec:model} for some representative cases, before presenting a simple analytical prescription. All models presented are at 5 AU with ISM grain opacity, with $M_e=M_c$. {Note that in general, since $E_G$ and $U$ have opposite signs, the magnitude of the total energy $E_G+U$ can be significantly smaller than the magnitudes of $E_G$ or $U$ alone. We discuss this further through the lens of the Virial theorem in Appendix \ref{Virial}.}

The results of the calculation of the binding energy are shown in Table \ref{tablenummimm}. 
For our nominal case (30 $M_\oplus$ core, $\gamma = 1.4$, and v$_\mathrm{imp}$ = v$_\mathrm{esc}$), the minimal impactor mass is 1.5 $M_\oplus$. It decreases mildly to 1.15 $M_\oplus$ for a 20 $M_\oplus$ core. In both of these cases, this mass is higher than those found for conditions 1 and 2, and are hence the minimal impactors masses needed to fully reset the corresponding pre-runaway envelope. 
For $\gamma = 1.25$, the masses are lower with 0.43 and 0.38 $M_\oplus$ for 30 and 20 $M_\oplus$ cores, respectively. In these cases, condition 1 is more stringent, although the masses found for condition 1 are also close to (but higher than) those needed to satisfy conditions 2 and 3. 

We note that in formulating condition 2 above (and in the previous sections), when calculating the core's post-collision temperature we assumed that the gravitational binding energy of the hydrodynamically lost fraction of the envelope (pre Eddington winds phase) is negligible compared to the kinetic energy of the impactor. This approximation is satisfied automatically when condition 3 is met. 


We now formulate a simple analytical prescription to calculate the minimal impactor mass needed, as given by condition 3 above. We start by defining $E_G^{a1}$ and $E_G^{a2}$ (evaluated in table \ref{tablenummimm}) to be
\begin{equation}\label{eq:EGa1}
E_G^{a1} = -G\frac{M_c M_e}{R_\mathrm{RCB}},
\end{equation}
and
\begin{equation}\label{eq:EGa2}
E_G^{a2} = -G\frac{M_c M_e}{R_{c}},
\end{equation}
where both are approximations to the gravitational binding energy of the envelope, but applicable under different conditions. For an adiabatic index $\gamma > $ 4/3, the envelope mass is concentrated closer to {the radiative-convective boundary (RCB), and hence we use the radiative-convective boundary radius $R_\mathrm{RCB}$ in the denominator.} For $\gamma \leq $ 4/3, the envelope is concentrated closer to the core, and hence the denominator is $R_c$ \citep{lee2}. We also define the approximate envelope's internal energy
\begin{equation}
U^{a} = C_V \bar{T} M_e,  
\end{equation}
where $\bar{T}$ is also the temperature of either the RCB, or the base of the envelope, depending on $\gamma$. Equating the impactor's kinetic energy with the envelope's total energy (|$E_G^{a}$ + $U^{a}$| ), for v$_\mathrm{imp}$ = v$_\mathrm{esc}$ we get the impactor's minimum mass as
\begin{equation}
M_{\rm imp} > \frac{M_{c}  M_{e}}{M_c + M_e}\frac{R_{c}}{R} - \frac{C_{V,e} \bar{T} M_e R_{c}}{M_c + M_e},
\label{eq:mimp}
\end{equation}
where $R$ is either $R_\mathrm{RCB}$ or $R_c$, depending on $\gamma$.

This equation however can be different from the numerical solution by a factor $1.5$--$3$ depending on the parameters, even if the correct $R_\mathrm{RCB}$ or $R_c$ was used. This is simply because, as can be seen in table \ref{tablenummimm}, the envelope's analytically calculated energy terms are different from the numerical values by factors of $1.5$--$3$. This is not surprising since equations (\ref{eq:EGa1}) and (\ref{eq:EGa2}) are missing order unity prefactors that depend on the density profile of the envelope. It is however useful as a starting point to estimate the order of magnitude minimal impactor mass, and can be calibrated using fudge factors to match the numerical simulations. 


\section{Post-collision cooling and reaccretion of the envelope}
\label{sec:reaccretion}

Once the core has cooled down to below the disc's local entropy, an atmospheric radiative zone can develop, allowing the envelope to cool radiatively and contract. The giant collision, however, could have deposited a large amount of solids in the envelope, leading to a higher metallicity envelope than the pre-collision counterpart. In the innermost regions of the envelope, the temperature is large enough to vaporize the silicate particles, possibly stopping these particles from being reaccreted onto the core and increasing the local gas phase opacity. Although a large fraction of the heavy-elemental debris may be lost with the initially-escaping super-Eddington wind to the disc region beyond the core's Hill radius, we assume it remains nearby and contaminates the ambient disc gas.  This metal-rich reservoir provides the gas supply for the subsequent re-accretion. 

In this section, we use a 1D atmospheric cooling-accretion model to calculate to what extent the envelope is able to be reaccreted following the giant impact, and thereby constrain the conditions under which we obtain an intermediate-mass giant planet ($\sim 50\ M_\oplus$). We first present details of the model (\S \ref{sec:model}), and then discuss how we treat the effects of heavy elements (\S \ref{sec:silicates}), in particular by calculating the wet adiabat that includes the effects of condensation of silicates. In \S \ref{sec:parameters}, we investigate the effects of the different model parameters through a grid search, to understand the range of possible outcomes of the post-collision evolution.

\label{sec:envelopeaccretion}
\subsection{Numerical model of cooling and accretion}
\label{sec:model}

The model is similar to the one explored in \cite{alidib}. We solve the classic interior structure equations for a constant luminosity envelope,
\begin{equation}
\label{eqsmain}
    \begin{aligned}
\frac{\mathrm{d} m}{\mathrm{~d} r} &=4 \pi r^{2} \rho \\
\frac{\mathrm{d} P}{\mathrm{~d} r} &=-\frac{G (M_{c} + m(r))}{r^{2}} \rho \\
\frac{\mathrm{d} T}{\mathrm{~d} r} &=\nabla \frac{T}{P} \frac{\mathrm{d} P}{\mathrm{~d} r}\\
P&=\rho k_{\mathrm{B}} T / \mu,
\end{aligned}
\end{equation}
where $m(r)$ is the envelope mass contained within radius $r$, $P$, $T$, and $\rho$ are the pressure, temperature, and density respectively, and $\mu$ is the mean molecular 
weight (reflecting the silicate content of the atmosphere). We include heat transport by radiative diffusion and convection by choosing the temperature gradient as
\begin{equation}
    \nabla=\min \left(\nabla_{\mathrm{ad}}, \nabla_{\mathrm{rad}}\right)
\end{equation}
where $\nabla_{\mathrm{ad}}\equiv\left(d\ln T/d\ln P\right)_{\mathrm{ad}}$ is the adiabatic gradient, and $\nabla_{\mathrm{rad}}$ is the radiative gradient
\begin{equation}
\label{deltarad}
    \nabla_{\mathrm{rad}} \equiv \frac{3 \kappa P}{64 \pi G (M_{c}+m) \sigma T^{4}} L
\end{equation}
with the atmospheric opacity $\kappa$ and luminosity $L$. We discuss our calculation of the adiabatic gradient and the distribution of silicates in the atmosphere in \S \ref{sec:silicates}.

We take into account contributions to the opacity from gas, grains, and H$^-$, writing $\kappa=\kappa_g + \kappa_d + \kappa_{H^-}$. We use an alternative opacity prescription from our previous paper \citep{alidib}, where we used \cite{belllin1994}, to allow us to explicitly include the gas and grain metallicities. With a solar value for metallicity $Z=0.02$, we find good agreement with \cite{belllin1994} for a wide range of temperature and density.

The gas opacity $\kappa_g$ is calculated using {equations (3)--(5) of \cite{freedman}}. This 
prescription includes an explicit dependence on the gas metallicity, and is thus suitable for a metal-enriched atmosphere. We emphasize that the metallicity here is the normalized \textit{gas phase} metallicity that is assumed to be solar (corresponding to the parameter ${met}=1$ in \citealt{freedman}) for the dry adiabat case (explained below), and is calculated consistently from the silicate vapor saturation density for the wet adiabat case. As the \cite{freedman} gas opacities were calculated for temperatures up to 4000 K, we extrapolate beyond this point. The H$^-$ opacity is taken from \cite{ferguson} and \cite{lee2}, 
\begin{equation}
    \begin{aligned}
\kappa\left(\mathrm{H}^{-}\right) \simeq & 3 \times 10^{-2} \mathrm{~cm}^{2} \mathrm{~g}^{-1}\left(\frac{\rho}{10^{-4} \mathrm{~g} \mathrm{~cm}^{-3}}\right)^{0.5} \\
& \times\left(\frac{T}{2500 \mathrm{~K}}\right)^{7.5}\left(\frac{Z}{0.02}\right).
\end{aligned}
\end{equation}

The  grain opacity is derived in \cite{alidibthomp}, and taken to be the minimum of twice the geometric opacity and the Rosseland mean small-grain opacity. We define
\begin{equation}\label{eq:kapgr}
\kappa_d = \kappa_{\rm geom}Q,
\end{equation} 
where $\kappa_{\rm geom} = 3Z_{gr}/4\rho_s a_d$ for spherical grains of radius $a_d$ (assumed 2.3$\times$10$^{-4}$ cm, value chosen for our ISM opacity case to be very close to the equivalent case using \citealt{belllin1994} and \citealt{ferguson}), material density $\rho_s$ (assumed 3.2 g/cm$^3$),
and mass fraction $Z_{gr}$, and
\begin{equation}\label{eq:Q}
  Q = {\rm min}(2, 0.35 \times 2\pi a_d / \lambda_{\rm max}), 
\end{equation}
where the peak wavelength in the Planck function is $\lambda_{\rm max}(T) \equiv hc/4.95\,k_{\rm B}T\approx 2.9\times 10^{-4}\ {\rm cm}\ (T/1000\ {\rm K})^{-1}$. 
With our choice of $a_d$ and $\rho_s$, the geometric opacity is $\kappa_{\rm geom}=10.2\ {\rm cm^2\ g^{-1}}\ (Z_{gr}/0.01)$.
We treat the mass fraction of grains $Z_{gr}$ as a free parameter, and take it to be independent of the wet/dry state of the envelope so we can separately understand the effects of the grain opacity. We turn off grain opacity at high temperatures to account for the sublimation of silicates. To achieve a smooth transition between the grain and H$^-$ opacities, we take the dust sublimation temperature to be 2500K (although larger than typically assumed, our results are not sensitive to the dust sublimation temperature since the envelope is typically adiabatic at this point).

To calculate the time-evolution of the envelope, we follow the prescription of \cite{piso} as described in \cite{alidib}. We construct a series of envelope models with increasing mass by integrating equations (\ref{eqsmain})--(\ref{deltarad}) inwards from the disc (at $r=R_{\rm out}$) to the surface of the core. We include a simplified model of recycling of disc material into the Hill sphere \citep{Ormel2015} as described in \cite{alidib}. We assume that the region $R_{\rm adv}<r<R_{\rm out}$ into which the recycling flow penetrates follows the same adiabat as the disc (since the flow rapidly advects the disc gas inwards). We show models below with ($R_{\rm adv}=0.3 R_{\rm out}$) and without ($R_{\rm adv}=R_{\rm out}$) recycling included (where $R_{\rm out}$ is the outer boundary of our integration). For each integration, we iterate to find the envelope luminosity (assumed to be radially constant in the radiative zone) corresponding to the desired envelope mass. 

We then connect these snapshots in time using the simplified energy equation
\begin{equation}
\label{tacc}
\Delta t=\frac{-\Delta E}{\langle L\rangle}
\end{equation}
where $\Delta E$ is the difference in total energy between two subsequent snapshots, and $\langle L \rangle$ the average of their luminosities. The total energy of the envelope for each snapshot is the sum of its gravitational and internal energies $E=E_{\mathrm{G}}+U$, with 
\begin{equation}
\label{egeq}
    E_{\mathrm{G}}=-\int_{M_{\mathrm{c}}}^{M} \frac{G m}{r} \mathrm{~d} m
\end{equation}
and
\begin{equation}
\label{ueq}
U=\int_{M_{\mathrm{c}}}^{M} u \mathrm{~d} m
\end{equation}
where the internal energy is $u=C_{V} T=\left(\nabla_{\mathrm{ad}}^{-1}-1\right) k_BT/\mu$.
We set the upper limit of these energy integrals to the location of the innermost radiative-convective boundary.

{In the above equations we are however ignoring the core terms. In reality, the core can act as either a heat source or sink if its surface temperature is different than that of the inner envelope. In Appendix B we explore the evolution of the envelope's base temperature, and find that, irrespective of the core, it can either increase or decrease with decreasing luminosity depending on the boundary conditions. Therefore, both directions of the core-envelope heat exchange are possible. Numerically examining multiple cases however reveals this exchange to be only relevant for $\sim 10^3$--$10^4$ yr , with the core becoming energetically negligible afterwards.  }

\subsection{Equation of state and condensation of silicates}
\label{sec:silicates}

Giant collisions vaporize and deposit large silicate mass into the envelope, increasing the mean molecular weight of the gas, and possibly saturating it, forcing a rising parcel to undergo condensation and triggering moist convection. We account for this possibility self-consistently by calculating a wet adiabat in the saturated regions of the atmosphere. For comparison, we also carry out simulations under the assumption of a standard envelope on a dry adiabat. 

\begin{figure*}
\begin{centering}
\includegraphics[scale = 1]{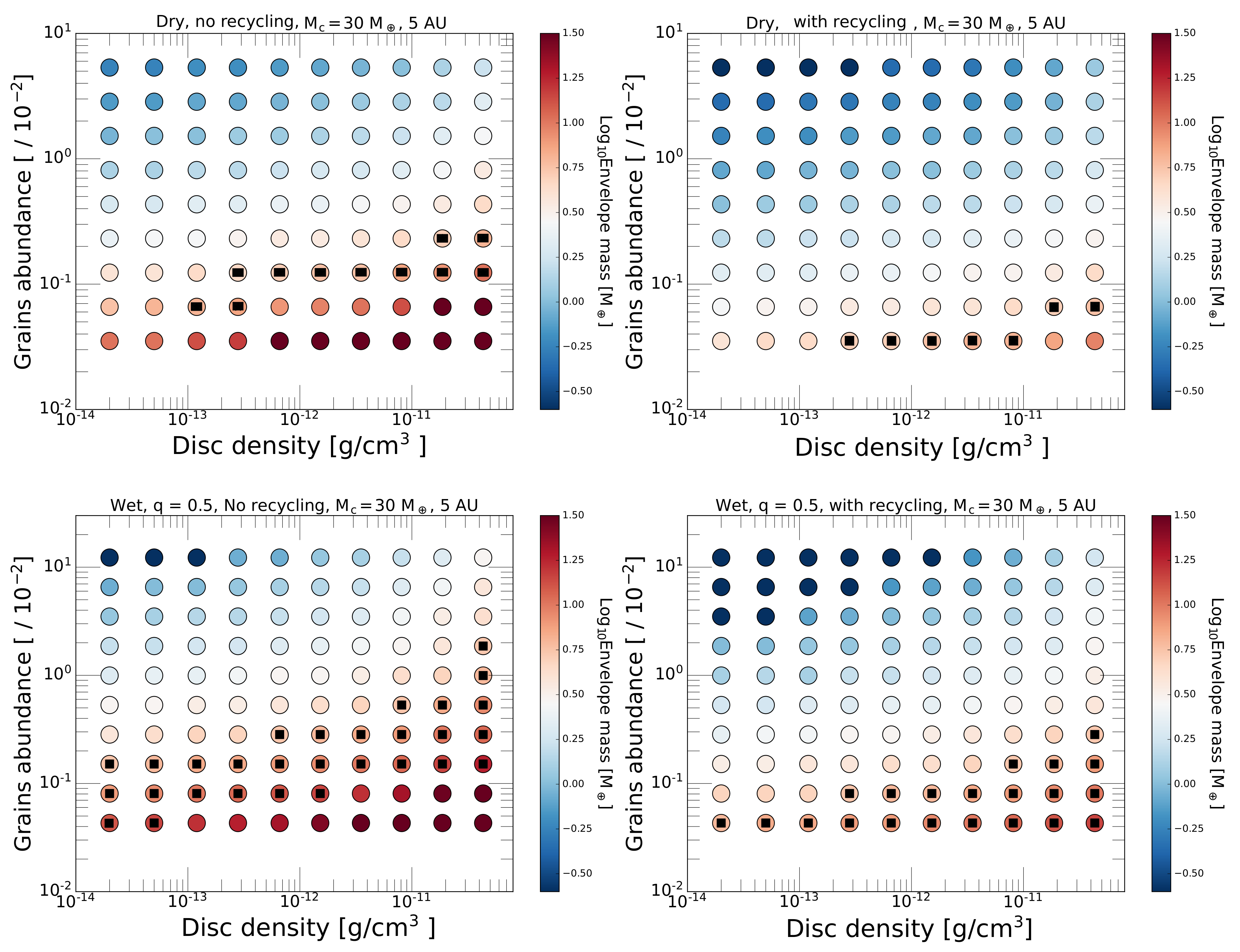}

\caption{The envelope mass of a 30 M$_\oplus$ core at 5 AU, for a cooling time $\sim$ 10$^5$ yr, for a variety of parameters. 
Top-row panels are on a dry adiabat, while bottow-row panels are on a wet 
adiabat and contain significant amounts of silicates. For both cases we show the total masses of the envelope, and this include all of the silicates in the wet adiabat case.
The left-column panels represent models without advective 
transport of disc entropy whereas the right-column panels corresponds to models which take into account of entropy advection.
Black markers represent cases with envelopes in the sub-Saturn regime and where the corresponding lower core-mass case 
(with $M_{c}=15 M_\oplus$) did not reach runaway accretion in $\sim 10^6$ yr. }   
    \label{fig:results1}
\end{centering}

\end{figure*}

In both cases, the opacity is controlled independently of the silicate content of the atmospheric gas, so that we can isolate the contributions from different physical effects. 
Having the opacity and silicate content be independent parameters also makes sense given the uncertainty in the size distribution of (undissolved) particles in the atmosphere. If these particles are large enough, then the atmosphere can be simultaneously be on a wet adiabat and have a low grain opacity. In the case of the dry adiabat, as this 
comparison model is an idealized control case, we do not try to make it fully self consistent, and thus allow for high grain opacity with low mean molecular weight. This approximation is only relevant on the higher end of our grain opacity values 
where their abundance is high enough to affect the mean molecular weight. 

For our envelope models that assume a dry adiabat, we take the composition of the atmosphere to be hydrogen and helium dominated, even for large grain opacity. For simplicity, we assume an adiabatic constant $\gamma=1.4$, { giving $\nabla_\mathrm{ad}=(\gamma-1)/\gamma=0.286$.}  In reality, under inner planetary envelope conditions, $\gamma$ can be as low as 1.2 due to hydrogen dissociation \citep{lee1,Piso2015}. In addition to decreasing the adiabatic temperature gradient, such low $\gamma$ values have indirect consequences as well, since the cooling rate of envelopes with $\gamma = 1.2$ has weaker dependence on the the boundary conditions than $\gamma = 1.4$ \citep{lee1,lee2}. However, since our focus is on the impact of heavy elements, we restrict our models to constant $\gamma=1.4$.

In the case of a wet adiabat, we start by assuming a (parametric) fixed silicate mass mixing ratio \begin{equation}  
q \equiv \frac{\rho_{Si,\mathrm{tot}}}{\rho_{Si,\mathrm{tot}} + \rho_{H_2}},
\end{equation}
where $\rho_{Si,\mathrm{tot}}$ is the total silicate density, then calculate the saturation mixing ratio $q_s$. For the region in the envelope where $q > q_s$, we assume the wet adiabatic gradient following \cite{leconte},
\begin{equation}
    \nabla_{\mathrm{ad}}^{\star} \equiv \frac{R}{\mu c_{p}}\left(1+\frac{q_{\mathrm{s}} M_{\mathrm{H}} \ell_{\mathrm{sil}}}
    {(1-q_{\mathrm{s}}) R T}\right) 
    /\left(1+\frac{q_{\mathrm{s}} \ell_{\mathrm{sil}} \gamma_{\mathrm{s}}}
    {(1-q_{\mathrm{s}})c_{p} T} \right),
\label{eq:nablawet}
\end{equation}
where the sublimation energy $\ell_{\mathrm{sil}}=1.6 \times 10^{11}$ erg $\mathrm{g}^{-1}$ \citep{Krieger} ,
\begin{equation} 
q_s \equiv \frac{\rho_{Si,\mathrm{sat}}}{\rho_{Si,\mathrm{sat}}+\rho_{H_2}},
\label{eq:qs}
\end{equation}
with $\rho_{Si,\mathrm{sat}}$ the saturated silicate vapour density, $M_{\mathrm{H}}$ and $M_{\mathrm{Si}}$ the molar masses 
of hydrogen and silicates, and
\begin{equation}
\left.\gamma_{\mathrm{s}} \equiv \frac{\partial \ln q_{\mathrm{s}}}{\partial \ln T}\right|_{p}=\left(1-\varpi q_{\mathrm{s}}\right) \frac{M_{\mathrm{v}} \ell_{\mathrm{sil}}}{R T} 
\end{equation}
with
\begin{equation}
    \varpi \equiv  \left( { M_{\mathrm{H}}-M_{\mathrm{Si}} \over  M_{\mathrm{H}} } \right).
\end{equation}

The silicate vapor density at saturation, and thus $q_{\mathrm{s}}$, is calculated as follows.
We first solve equations (\ref{eqsmain})--(\ref{deltarad}) assuming a dry adiabat. This allows us to obtain a first estimate for the temperature and H$_2$ gas density $\rho_{H_2}$ profiles. 
We then calculate the saturation pressure of \cite{Krieger} as
\begin{equation}
P_{\text {Si }}^\mathrm{sat}(T)=3.2 \times 10^{14}\ {\rm erg\ cm^{-3}}\  e^{-\left(6 \times 10^{4} \mathrm{~K}\right) / T} 
\end{equation}
followed by the saturation Si vapor density
\begin{equation}
    \rho_{Si,\mathrm{sat}} = \frac{P_{\text {Si }}^{\mathrm{sat}} \mu_{Si}}{k_\beta T}.
\end{equation}
With $\rho_{Si,\mathrm{sat}}$ and $\rho_H$ in hand, we now calculate the $q_s$ profile of the envelope. Finally, with $q_s$ known we reintegrate equations (\ref{eqsmain})--(\ref{deltarad})  but for the wet adiabat case where the adiabatic gradient is solved consistently for each radial step as long as the parametric $q$ is larger than the calculated $q_s$. A dry adiabat is assumed in the inner envelope where this is not satisfied.  

\begin{figure*}
\begin{centering}
        \includegraphics[scale=0.27]{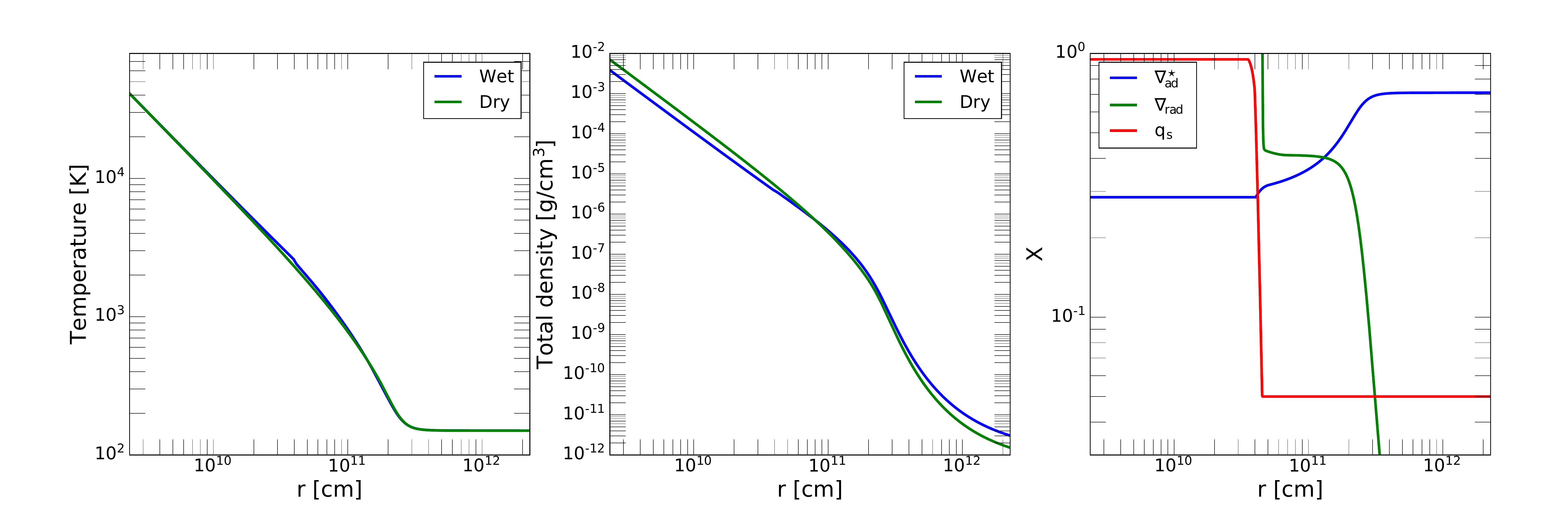}
   \caption{Left and center: The temperature and density profiles in the envelope for wet and dry adiabat cases. Note that the wet case's density profile is a factor of two higher than that of the dry case around the outer boundary because we are imposing wet $q=0.5$ while maintaining the disc's density to be constant. Right: the radiative and moist adiabatic gradients (Eq. \ref{eq:nablawet}) of our nominal case, in addition to the saturation mass mixing ratio of silicate $q_s$ (Eq. \ref{eq:qs}). All curves are for $t\sim 10^5$ yr, disc density = 1.5$\times 10^{-12}$, and grain opacity of 1.8 $\times$ ISM. The wet adiabat curves are for M$_{e}\sim $ 2.04 M$_{\oplus}$ and $L$=1.2$\times$ 10$^{26}$ erg/s, while the dry adiabat curves are for  M$_{e}\sim $ 2.42 M$_{\oplus}$ and $L$=9.18$\times$ 10$^{25}$ erg/s.  }
    \label{fig:compwd}
    \end{centering}
\end{figure*}
\subsection{Results and exploration of parameter space}
\label{sec:parameters}

We now investigate the circumstances under which the core is able to accrete enough gas following the giant collision to produce a sub-Saturn mass planet. We also check whether the cores that collided would have remained sub-critical (avoiding runaway accretion) in the protoplanetary disc before the collision occurred.

The results of our numerical model are presented in Fig.~\ref{fig:results1}. These plots show the envelope mass around a 30 M$_{\oplus}$ core after $10^5$ yr of accretion. We choose a timescale of $10^5\ {\rm yrs}$ because this is the characteristic gas depletion timescale for transitional discs.
In these models, we consider a core of 30 M$_{\oplus}$ located at 5 AU around a solar mass star, where the disc temperature is 150 K. 
Each panel shows the envelope mass as a function of disc density $\rho_d$, and the mass fraction of grains $Z_{\rm gr}$ (the parameter that determines the grain opacity). We show four panels for different choices of wet or dry adiabats, and with or without envelope recycling. We note, for reference, that the minimal mass solar nebula has a density at 5 AU of $3\times 10^{-11}$ g/cm$^3$, with a passive disc temperature (not including viscous heating) of 60 K. 
We normalize the grain abundance on the y axis to the nominal solar value of $Z_{gr} = 10^{-2}$.

For the wet adiabat cases, we assume a constant silicate mixing ratio of $q=0.5$. This implies that the protoplanet is
accreting a silicate-rich envelope with equal masses of hydrogen and silicates. Such silicate-enriched gas in the 
feeding zone of the protoplanet is expected after a giant collision, near the snow line, or the inner edge of the
MRI-inactive zones \citep{kretke2007, kretke2009}. Note that the dry adiabat case corresponds to $q\ll 1$. Therefore, 
the wet and dry adiabat cases can be seen as limiting cases for the role of silicates in the envelope.

For each set of parameters, we also run corresponding simulations with lower mass cores, $M_{c}= 7.5 M_\oplus$ and 
15 M$_{\oplus}$. In Fig.~\ref{fig:results1}, we indicate with black rectangles the cases that satisfy both of the
following conditions: 1) the corresponding simulations with lower-mass cores ($7.5$ and $15 M_\oplus$) do not undergo
runaway accretion in $\sim$ 10$^6$ yr (comparable to the age of protostellar-disc bearing T Tauri stars), and 2) the final
envelope mass at 10$^5$ yr around the 30 M$_{\oplus}$ cores is $10$--$20 M_{\oplus}$ (corresponding to an asymptotic
planet-to-star mass ratio $q\sim 10^{-4}$). Therefore the black rectangles represent cases where low-mass cores remain
sub-critical (avoiding runaway accretion) during the main evolutionary phase of protoplanetary discs and accrete
only enough envelope during the transitional disc phase to produce a sub-Saturn planet following the merger of two
equal-mass (with $M_c = 15 M_\oplus$ each) cores.

\subsubsection{Models with a dry adiabat}
The top panels of Fig.~\ref{fig:results1} show results for models with dry adiabats. Without recycling (top-left panel of Fig.~\ref{fig:results1}), there is a shallow ``diagonal'' in disc density--grain abundance space, extending from $(\rho_d, Z_{gr})\approx (10^{-13}$ g/cm$^3, 7\times 10^{-4})$ to $\approx (8\times 10^{-11}$ g/cm$^3, 3\times 10^{-3})$, in which the solutions satisfy both of the conditions for forming a sub-Saturn (black rectangles). 
Below this diagonal, either the 30 M$_{\oplus}$ core reaches runaway accretion in $10^5$ yr, or the lower mass cores reach runaway in $10^6$ yr. Above this diagonal, our simulations lead to low envelope masses ranging from few $M_\oplus$ to virtually no envelope at all. Comparing this case to the corresponding simulations with entropy advection (top-right panel of Fig.~\ref{fig:results1}), we find similar trends but with the CSS diagonal pushed down to $(\rho_d, Z_{gr})\approx (3\times 10^{-13}$ g/cm$^3$, $4\times 10^{-4})$ and $\approx (8\times 10^{-11}$ g/cm$^3$, $7\times 10^{-4})$. 
Entropy advection hence has the effect at 5 AU of increasing  cooling times by a factor of $\approx 4$, in agreement with our previous results \citep{alidib}.

\subsubsection{Models with a wet adiabat}

In the bottom panels of  Fig. \ref{fig:results1},   we show the \textit{total} envelope mass, including both hydrogen and silicates. 
As we are imposing a silicates mass fraction of 0.5, there is by design equal amounts of hydrogen and silicates in the envelope. 

We start with the no-recycling case that leads to a wider CSS solutions diagonal compared to the  analogous dry envelope model, with both models centering around the same density-opacity diagonal. 
Overall, we find that considering wet adiabats in the explored part of parameters space leads to only moderately different envelope masses compared to the dry cases.

To better understand the reasons, we plot in Fig. \ref{fig:compwd} the temperature and total density profiles for some typical wet and dry adiabat cases having the same core mass, disc density, and grain abundance Z$_{gr}$. We also plot the thermal gradients and silicates mixing ratio of the wet case. 
This plot shows only minor differences in the total density profiles between the two cases. Comparison between the radiative and moist adiabatic gradients in the wet case reveals three thermal regimes: 1) In the outer envelope, where the saturation pressure of hydrogen is very low leading to a very small \textit{saturation} 
silicates vapor mixing ratio q$_s$, the radiative gradient is an order of magnitude lower than the wet adiabatic gradient, 
leading to a typical radiative and isothermal outer atmosphere. 
2) In the mid-envelope, as the temperature is increasing, the radiative gradient becomes larger than the wet adiabatic 
gradient, creating a narrow wet adiabatic mid-atmosphere with steep density profile. 
3) In the inner envelope, as the temperature (and thus q$_s$) increases exponentially (making q$_s$ $>$ $q$ = 0.5), the envelope connects to a dry adiabat.

Therefore, for the wet adiabat case to diverge significantly from the dry adiabat case, either the 
ambient temperature needs to be significantly larger to increase the silicates saturation pressure in the outer envelope, or 
$q$ needs to be significantly larger than the adopted value of 0.5 to widen the wet mid-envelope, although leading to extremely metal-rich envelopes.  

Physically, multiple different processes contribute to the differences between the wet and dry adiabat cases. First, since a larger adiabatic gradient (as in for a wet adiabat) pushes the RCB inwards to higher pressures, and since  $t_{\mathrm{cool}} \propto P_{\mathrm{RCB}}^2$ \citep{piso}, the cooling-accretion time is larger in this case. 
This tendency can however be offset by the effects of the mean molecular weight, where higher envelope metallicity decreases the cooling time by increasing the temperature and therefore luminosity necessary to maintain hydrostatic equilibrium \citep{lee2}.


The lower right panel  of Fig.~\ref{fig:results1}
shows the case with both a wet adiabat and envelope recycling. This has again lower envelope masses than the equivalent non-recycling case. We find virtually no protoplanets reaching runaway accretion in any part of parameter space. Only envelopes with the lowest envelope grain abundances ($\sim$ 4-6 $\times$ 10$^{-4}$) can accrete enough mass to be in the CSS regime.

\begin{figure*}
\begin{centering}
        \includegraphics[scale=0.40]{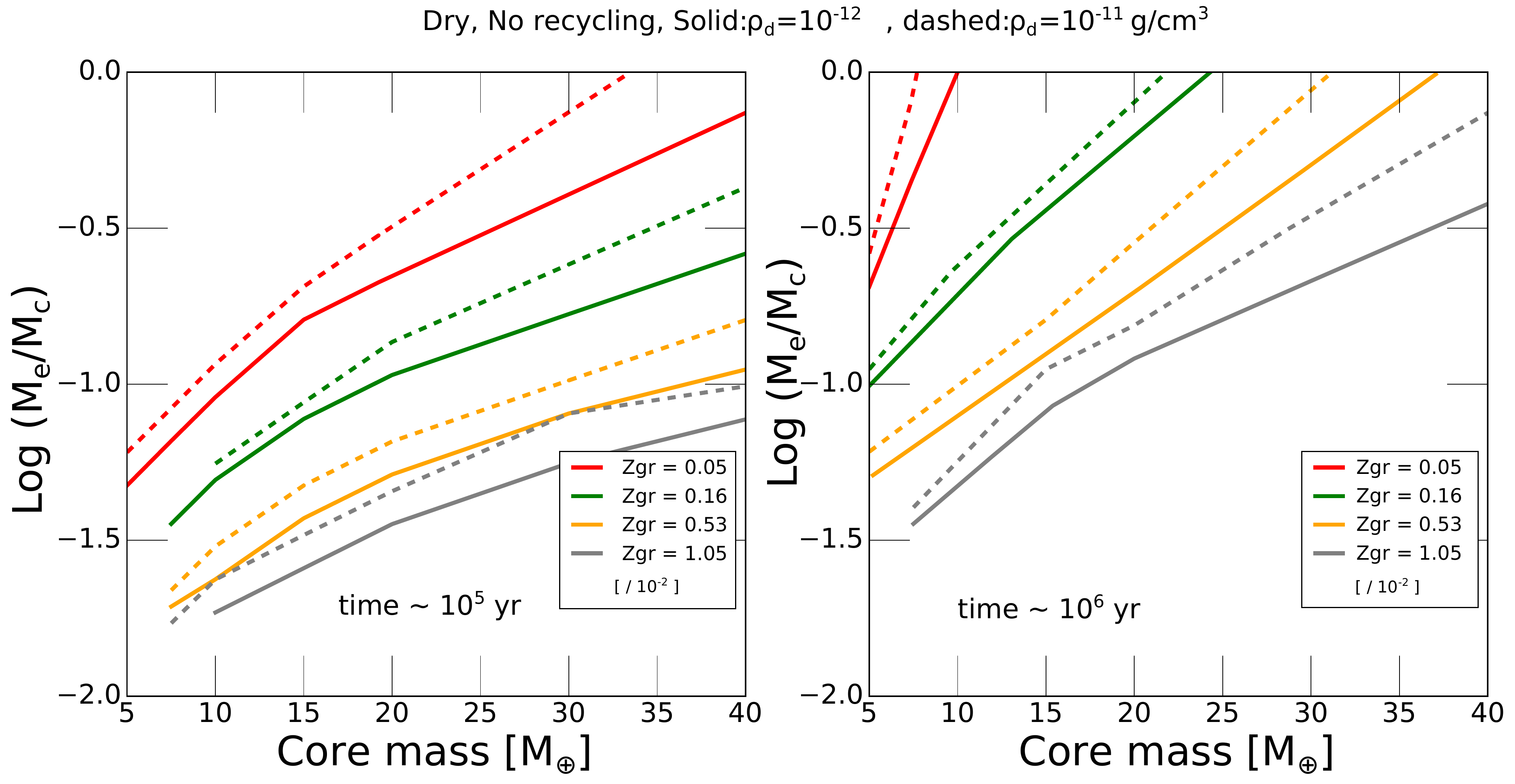}
        \includegraphics[scale=0.40]{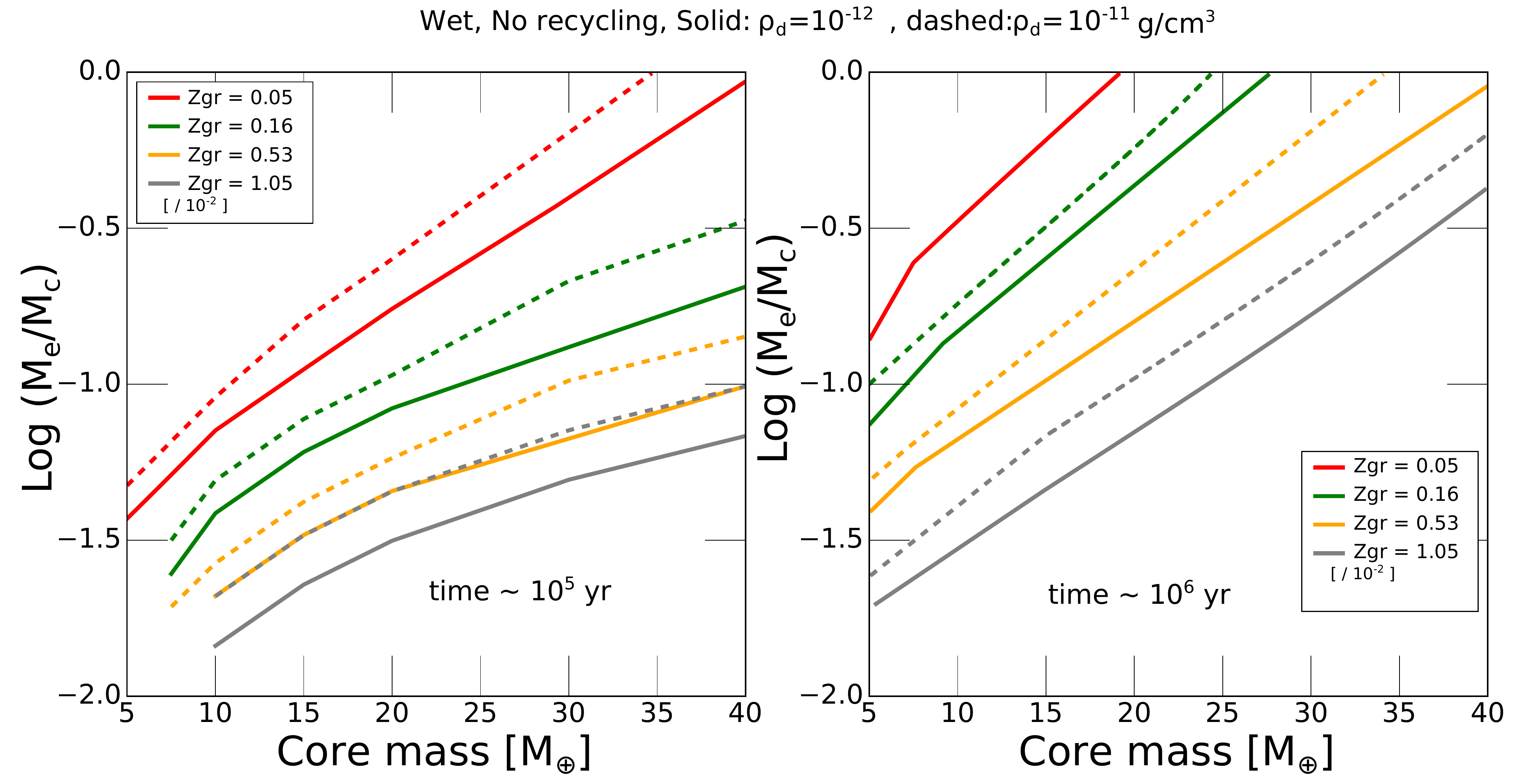}
   \caption{The envelope to core mass ratio for a wide range of core masses (5 to 40 $M_{\oplus}$) and grain opacities $Z_{gr}$ (5$\times$10$^{-4}$ to 10$^{-2}$). Solid and dashed lines represent different disc densities. Plotted grain abundances are all in units of solar metallicity. } 
    \label{fig:massgrid2}
    \end{centering}
\end{figure*}

\subsubsection{Effects of core mass}

\begin{figure*}
\begin{centering}
        \includegraphics[scale=0.40]{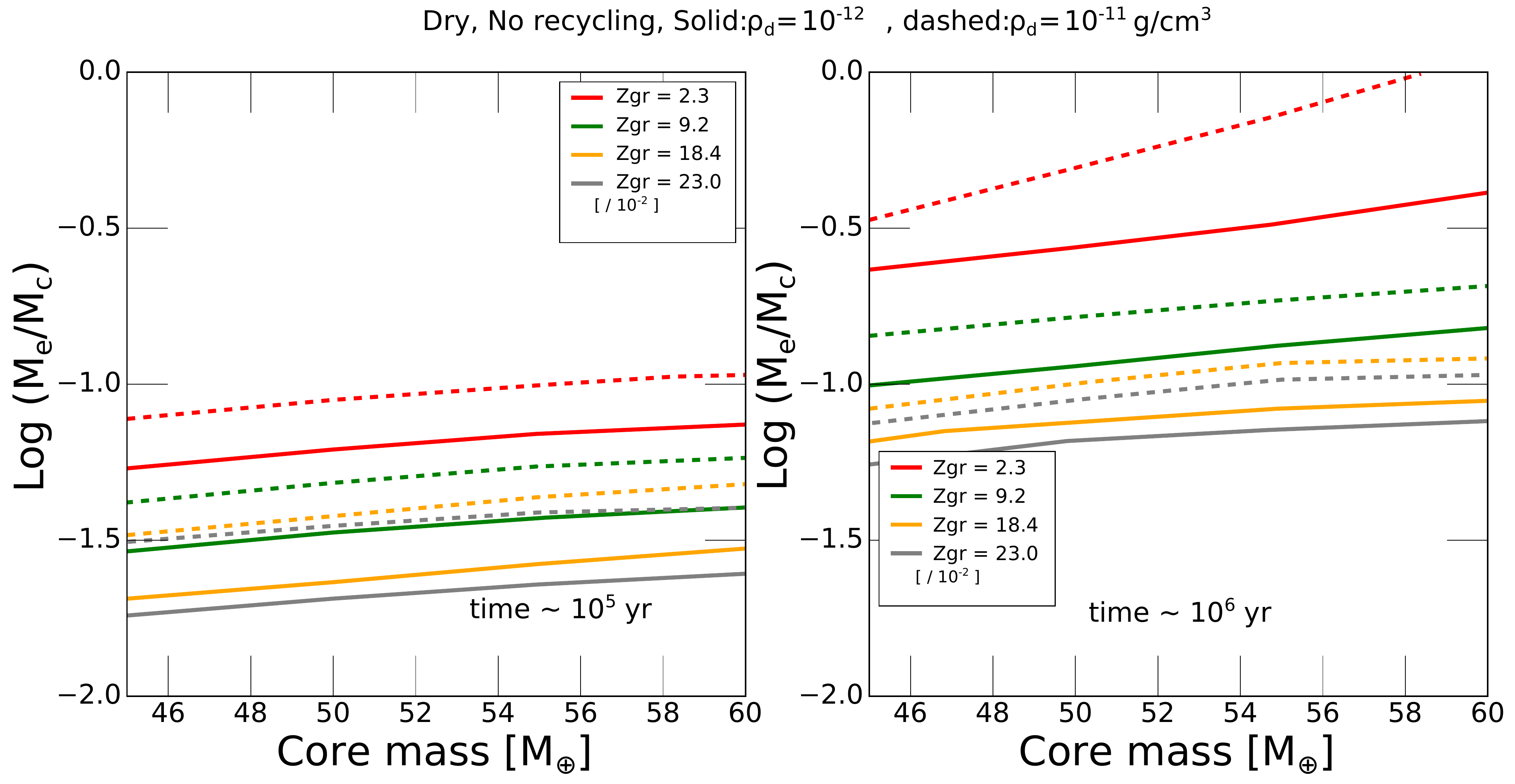}

        \includegraphics[scale=0.40]{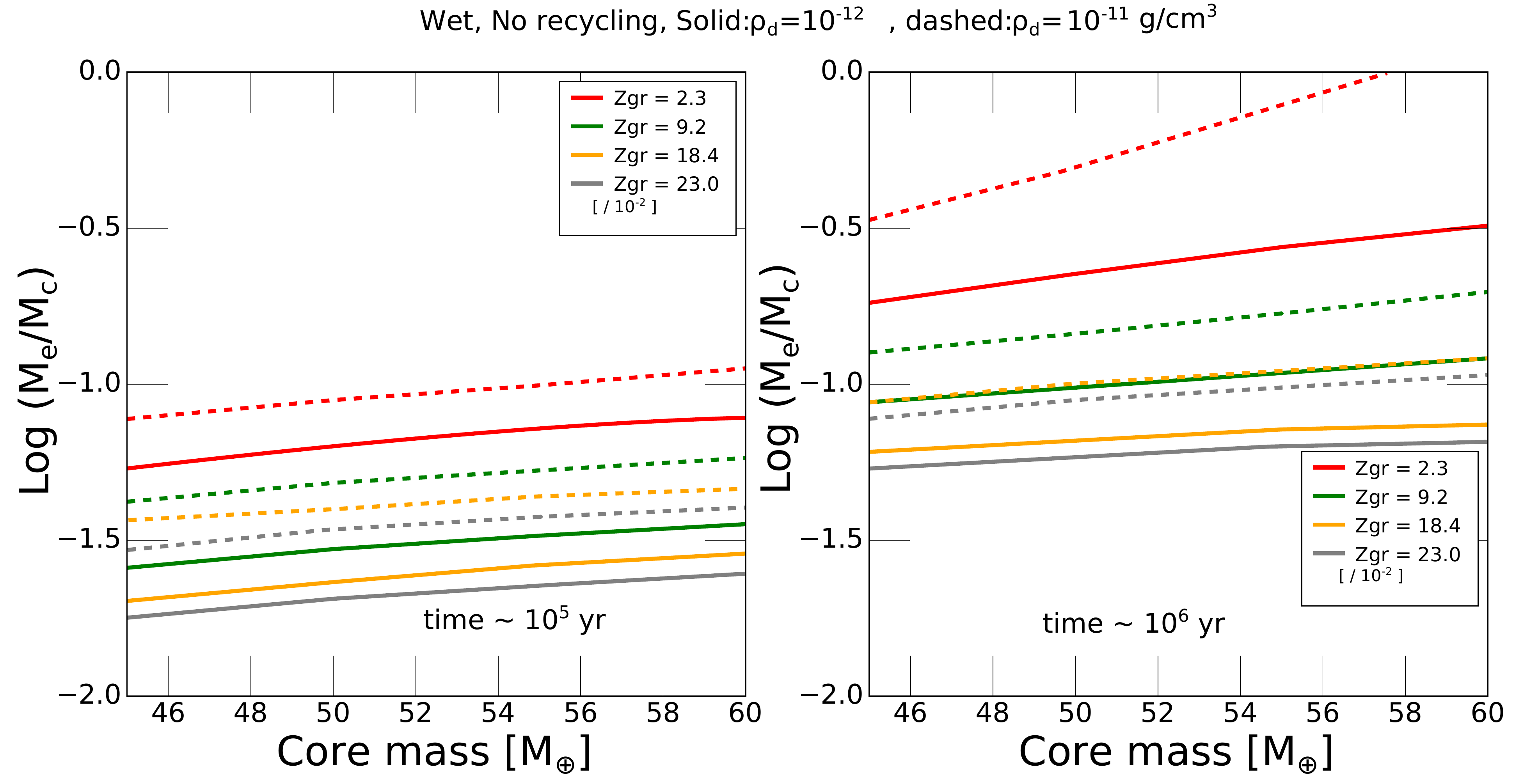}
   \caption{Same as \ref{fig:massgrid2}, but for higher core masses and higher grain opacities. }
    \label{fig:massgrid3}
    \end{centering}
\end{figure*}

In order to explore the effects of different core masses, we show, in Fig.~\ref{fig:massgrid2}, the envelope-to-core mass
ratio at $\sim$ 10$^5$ yr (left panel) and $\sim$ 10$^6$ yr (right panel) as a function of the core mass, metallicity, and gas density in the disc.  The goal in this parameter study is to investigate further whether cores can indeed grow into $\sim$ 30-40 M$_{\oplus}$ via giant mergers in relatively low-density transition discs (on the time scale of $10^5$ yr) without any of their ``building-block'' progenitors undergo runaway accretion in the relatively high-density protostellar disc (on the time scale of $10^6$ yr).
For both the dry and wet envelope cases, and $\rho_d = 10^{-12}$ and $10^{-11}$ g/cm$^3$ ({\color {black} i.e. $\sim 0.1-1$ that of the minimum mass nebula at 5AU}), we find that all core mass series reach runaway within $\sim$ 10$^6$ yr for grain abundances $\lesssim 0.5 \times $ ISM. This result is consistent with Fig. \ref{fig:results1}, and it reveals a rough minimum envelope grain opacity value needed to form CSS cores via giant mergers. 

In Fig. \ref{fig:massgrid3}, we also investigate the effects of core masses but for values in the 40-60 M$_{\oplus}$ range and for mostly supersolar grain opacities. These cases reveal that while a $M_{c}$=55-60 M$_{\oplus}$ core with a solar opacity envelope and a disc density of $10^{-11}\ {\rm g\ cm^{-3}}$ does runaway in  $\sim$ 10$^6$ yr, envelopes around $M_{c}=40$--$60\ M_{\oplus}$ cores with supersolar opacities can indeed survive the protoplanetary disc phase while remaining below crossover mass.

\section{Formation of cold Sub-Saturns \& ice giants}
\label{sec:icegiants}

We now discuss the implication of our results in \S\ref{sec:giantimpact} \& \S\ref{sec:reaccretion} in the context of a plausible formation channel for cold sub-Saturn (CSS) planets as qualitatively portrayed in Fig.~\ref{fig:cartoon}.  The basic initial condition we assume is the prior emergence of multiple ice-giant progenitors with sub-critical masses such that they elude runaway gas accretion during the main evolutionary phase of their natal discs.

These protoplanets are prone to undergo collisions during the transitional stages of their natal discs when the gas density substantially decreases on the timescale of $\sim 10^5$ years. We have already demonstrated (in \S\ref{sec:giantimpact}) that such collisions can lead to the ejection of the planets' original envelope while their cores coalesce with little mass loss from their merged cores (Fig. \ref{fig:planets}). In this section, we examine the necessary condition for these merged cores to re-accrete a modest-mass envelope without undergoing runaway accretion and transforming into gas giants before their natal discs are severely depleted during the transitional epoch.

\subsection{Formation pathways of cold sub-Saturn-mass planets}

\begin{figure*}
\begin{centering}
        \includegraphics[scale=1.1]{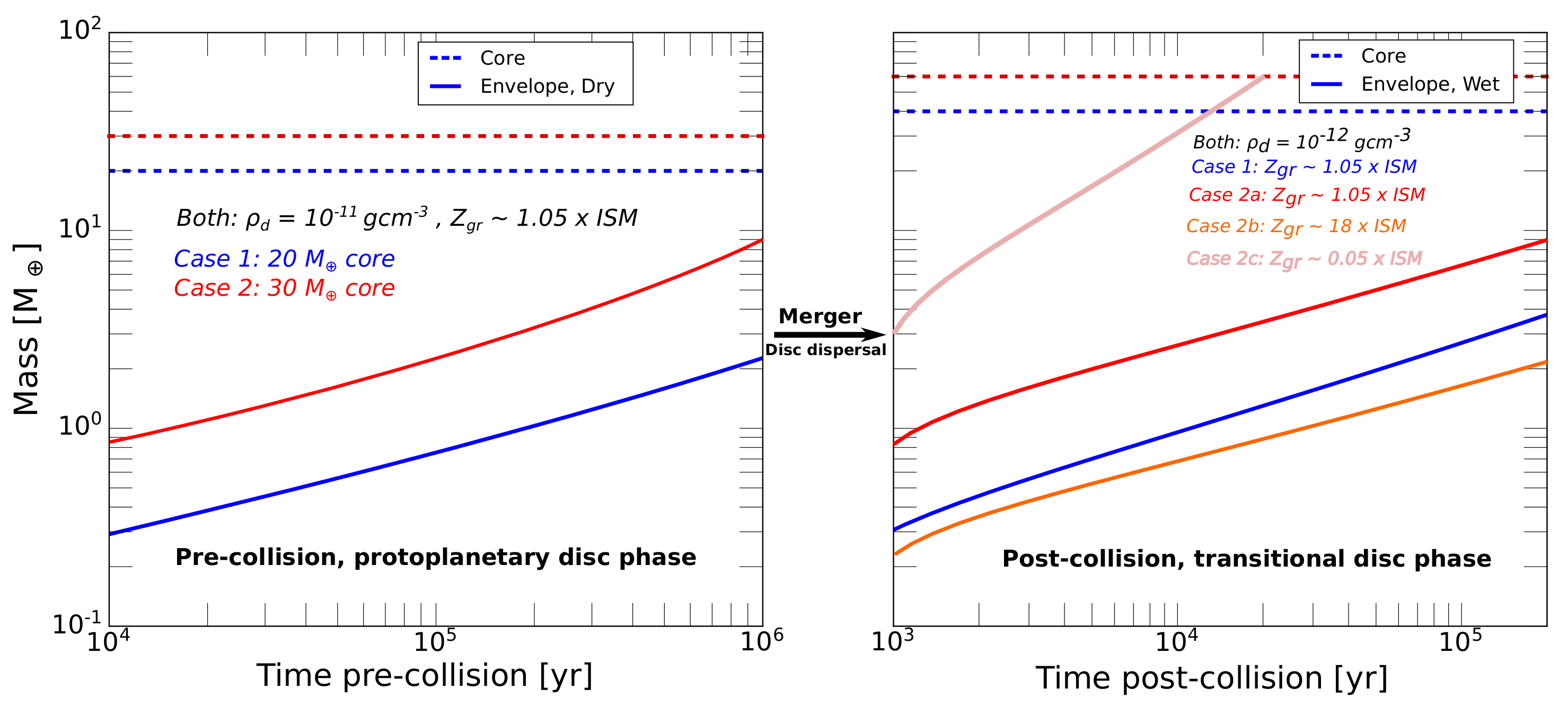}
   \caption{Blue lines: example for a cold sub-Saturn formation track, for Z$_{gr}$ = 1.05$\times$ ISM. A 20 M$_{\oplus}$ cores embedded at 5 AU in a protoplanetary disc accretes dry gas for 10$^6$ yr but remains sub-critical. During the initial dispersal phase of the disc, the protoplanet undergoes a giant collision and merger with a equally-massed protoplanet, resulting in a 40 M$_{\oplus}$ naked core. This new core undergoes a wet gas accretion phase in a lower density environment for $10^5$ yr, till the transitional disc disperses entirely.  Red lines: example for a cold sub-Saturn formation track but starting with two heaver cores (30 M$_{\oplus}$) that merge into a massive core accreting metal-rich gas. This case may result from the partial ejection of the core material into the disc after giant impacts, significantly raising ambient metallicity. Orange and pink lines: same as red, but for lower post-collision grain opacity. In all cases, solid and dashed lines represent the envelope and core masses, respectively.}
    \label{fig:massevolution}
    \end{centering}
\end{figure*}

\subsubsection{A monolithic-growth scenario for the progenitor cores}

We first consider a limiting case of a giant collision between a 30-$M_\oplus$ core and an impactor with sufficient mass to clear its progenitor's envelope without significantly increasing the mass of the merged core. 

With a modest (solar or slightly sub-solar) grain abundance, there exists a significant range of model parameters (with dry and wet envelopes, with and without entropy advection) that allows progenitors with a 30 M$_\oplus$ prior-formed core to acquire a comparable-mass envelope in 1 Myr without the onset of runaway accretion (Fig. \ref{fig:results1}). Since these models are constructed for cores at 5 AU, the contribution from recycling has limited effects on their results. The main determining parameters are the core mass $M_c$ and grain abundance $Z_{gr}$ in the envelope. With a super-solar $Z_{gr}$, the magnitude of the critical core mass can match the mass range of CSS (Fig. \ref{fig:massgrid3}) with little or no post-collision gas accretion.

However, there are two potential issues associated with this essentially {\it monolithic} growth scenario in which the progenitors presumably acquired their core mass through pebble accretion and the contribution of giant impacts is limited to the clearing of the pre-collision envelope {\it without} any changes in the core mass.

First, with a range of relatively-high disc density and relatively-low grain abundance (the right bottom of each panel in Fig. \ref{fig:results1}), some cores are able to re-accrete within $10^5\ {\rm yr}$, comparable-mass envelopes such that their total mass becomes comparable to that of CSS. However, these boundary conditions also introduce a conundrum because we would expect such large cores to achieve runaway envelope growth during the main evolutionary stage of their natal disc (on a timescale $\gtrsim 10^6$ yr), before the timely onset of disc-gas depletion or a major sustained enhancement in the envelopes' grain abundance (e.g.~see the right hand panel of Fig.~\ref{fig:massgrid2}).

Second, the minimum mass required for a core to accrete a comparable-mass envelope within $10^5\ {\rm y}$ is $\sim 30 M_\oplus$ (\S\ref{sec:parameters}).  A potential obstacle for this required initial condition is pebble isolation, which occurs when the protoplanets acquire sufficient mass to perturb the gas surface density distribution and induce the opening of narrow gaps. In principle, local gas pressure maximum stalls pebbles' orbital migration and quenches their supply to embedded cores \citep{ormel2010, ormel2017, Lambrechts2012, Lambrechts2014}.  The accumulation of pebbles at the migration barriers also increases their collision frequency and enhance their fragmentation rates. A continual flux of sub-millimeter grains bypasses the migration traps and reaches the proximity of the cores and elevate the opacity to super solar values. These effects alone may be adequate to suppress the transition to runaway accretion and preserve relative massive cores albeit entropy advection between the envelope and the disc can further reduce the accretion rate
\citep{chen2020}. Sustained pebble supply also provides a mechanism for cores to acquire masses larger than the isolation mass. Nevertheless, these effects may not be adequate to promote formation of core with $M_c \gtrsim 30 M_\oplus$.

\subsubsection{A merger hypothesis for the progenitor cores}

In order to bypass these issues, we now take into account the contribution of giant impacts to the growth of core mass.  We define super-critical/sub-critical cores to be those which can or can not acquire equal mass envelopes (i.e. $M_e \gtrless M_c$) during the active phase (within 1 Myr) of protoplanetary discs (with $\rho_d = 10^{-11}$ g cm$^{-3}$).  
The coalescence of cores with comparable sub-critical ($M_c \lesssim 20-30 M_\oplus$) masses increases their combined mass by up to a factor of two (to super-critical values of $M_c \gtrsim 40 M_\oplus$, Fig.~\ref{fig:cartoon}). These building blocks can attain their modest core masses without having to bypass the pebble-isolation barrier. Moreover, they also avoid runaway accretion during the main evolutionary stage of their natal discs (Fig.~\ref{fig:massgrid2}).  Although the total mass of their coalesced cores is near the low-end of the mass range of CSS, the super-critical merger products jump start efficient gas accretion from their natal discs. The preferred occurrence of these merger events in rapidly depleting transitional discs also suppresses the possibility of runaway accretion and provides a natural limit on the planets' asymptotic mass (below that set by the gap-in-gaseous disc condition, see \S\ref{sec:introduction}). A detailed 
example is illustrated in Fig.~\ref{fig:massevolution}. 

The transition between CSS and gas giants would be stifled if $M_{e}/M_{c} \lesssim 1$
for both the progenitors (within $\sim 10^6$ yr) and their merger products (within $\sim 
10^5$ yr). These constraints limit the asymptotic mass of the emerging planets
to be no more than four times the critical mass for cores to acquire $M_{e} \simeq M_{c}$
during the main evolutionary phase of their natal discs ($\sim 10^6$ y).  Whether the planets' asymptotic mass is within the mass range of CSS depends on the grain abundance in the envelope. 
With solar grain abundance, a $\sim 30 M_\oplus$ core acquires $M_{e}/M_{c} \sim 0.2$ 
from the disc (with $\rho_d =10^{-11}$g cm$^{-3}$) after $\sim 10^6$ yr (right panels Fig. 
\ref{fig:massgrid2}).  If a giant impact produced a $\sim 60 M_\oplus$ core, it would 
acquire a comparable $M_{e}/M_{c}$ from a somewhat depleted $(\rho_d \lesssim 10^{-12}$ g cm$^{-3}$) transitional disc in $10^5$ yr (left panels Fig. \ref{fig:massgrid3}).  

These critical masses increase with the grain abundance. If a fraction of the cores of the colliding planets are returned to their natal discs or mixed into the retained envelope \citep{liu2015b, liu2019}, the merged cores would be surrounded by heavy-element-rich envelopes.  With twice or more solar grain abundance, a $\sim 45 M_\oplus$ core accretes an envelope with $M_{e}/M_{c} \lesssim 0.3$ during the main evolutionary phase of the disc (right panels Fig. \ref{fig:massgrid3}).  Its coalescing collision (during either the main and transitional phases) with a less massive (e.g. $M_c \sim 15 M_\oplus$) core can still lead to the emergence of a byproduct with asymptotic total mass comparable to those of the CSS. For the same asymptotic total mass, the ratio $M_{e}/M_{c}$ decreases with 
the grain abundance in the envelope (Fig. \ref{fig:massgrid3}). As these initial and 
boundary conditions depend not only on the stage of the disc's evolution but also on the 
stellar parameters, this range of possibilities imply that CSS planets have the potential 
to form with diverse core-envelope ratios in transition discs around a wide variety of stars.



\begin{figure*}
\begin{centering}
\subfloat{\includegraphics[width = 3.5in]{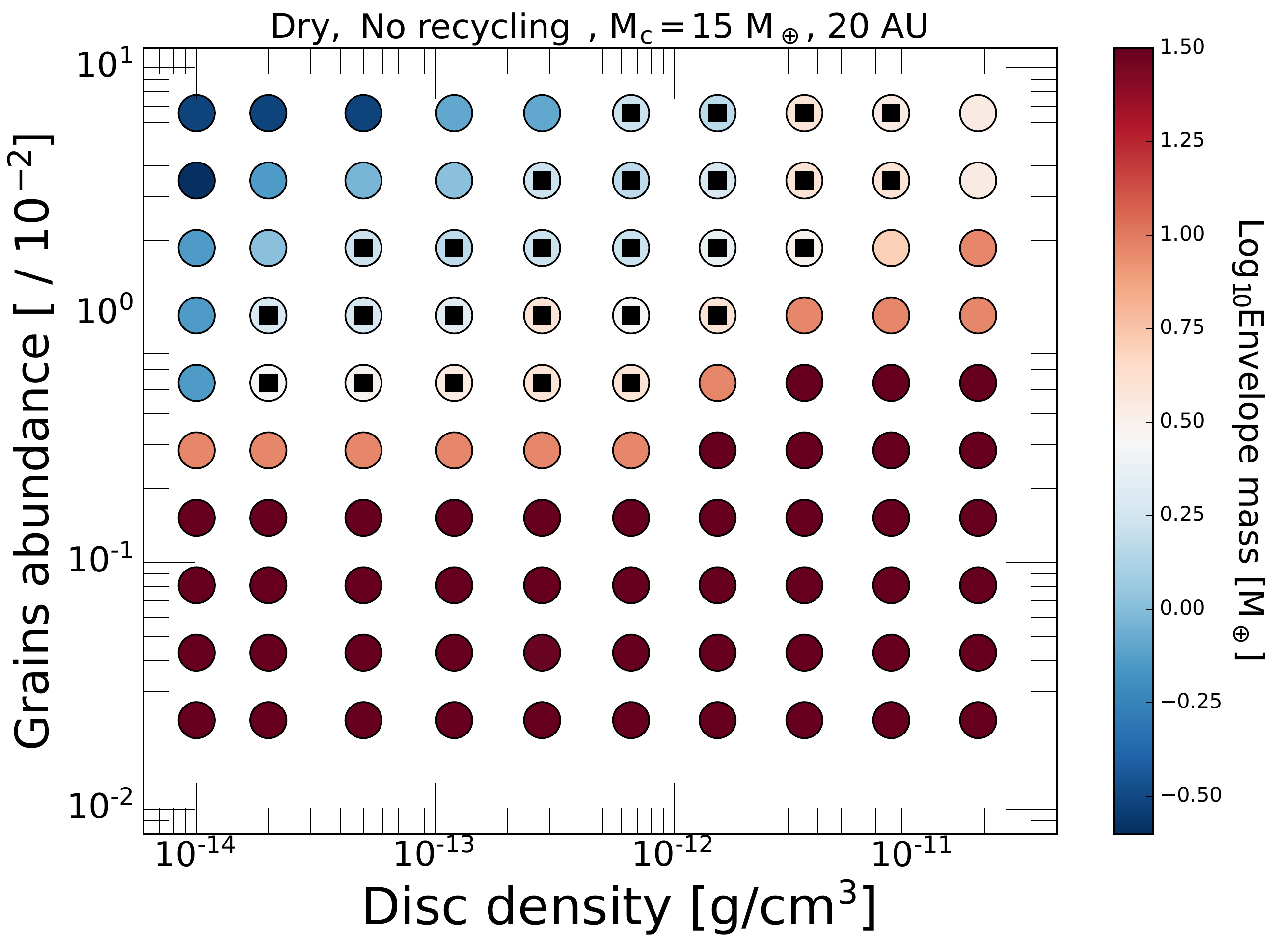}} 
\subfloat{\includegraphics[width = 3.5in]{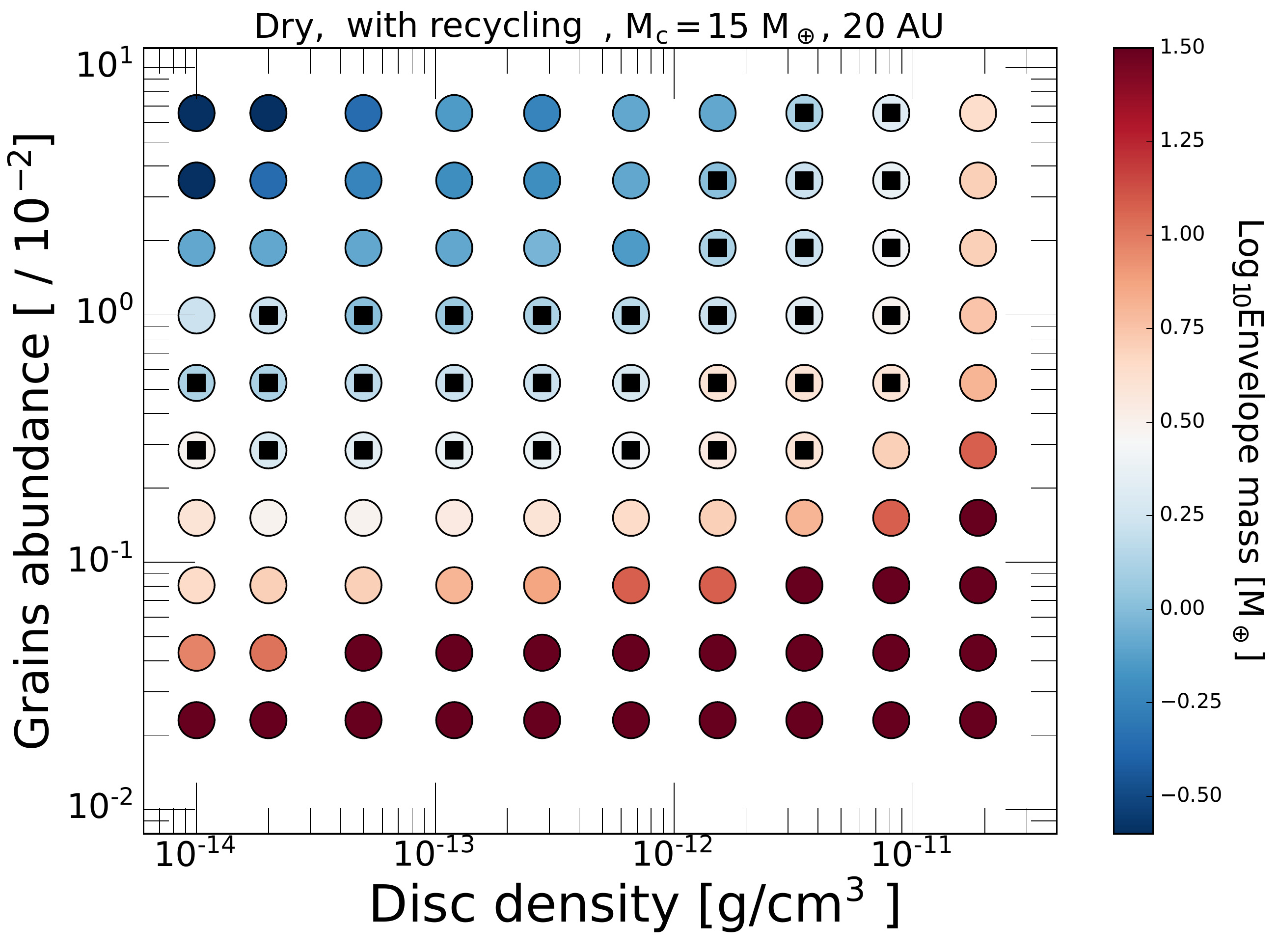}}\\
\subfloat{\includegraphics[width = 3.5in]{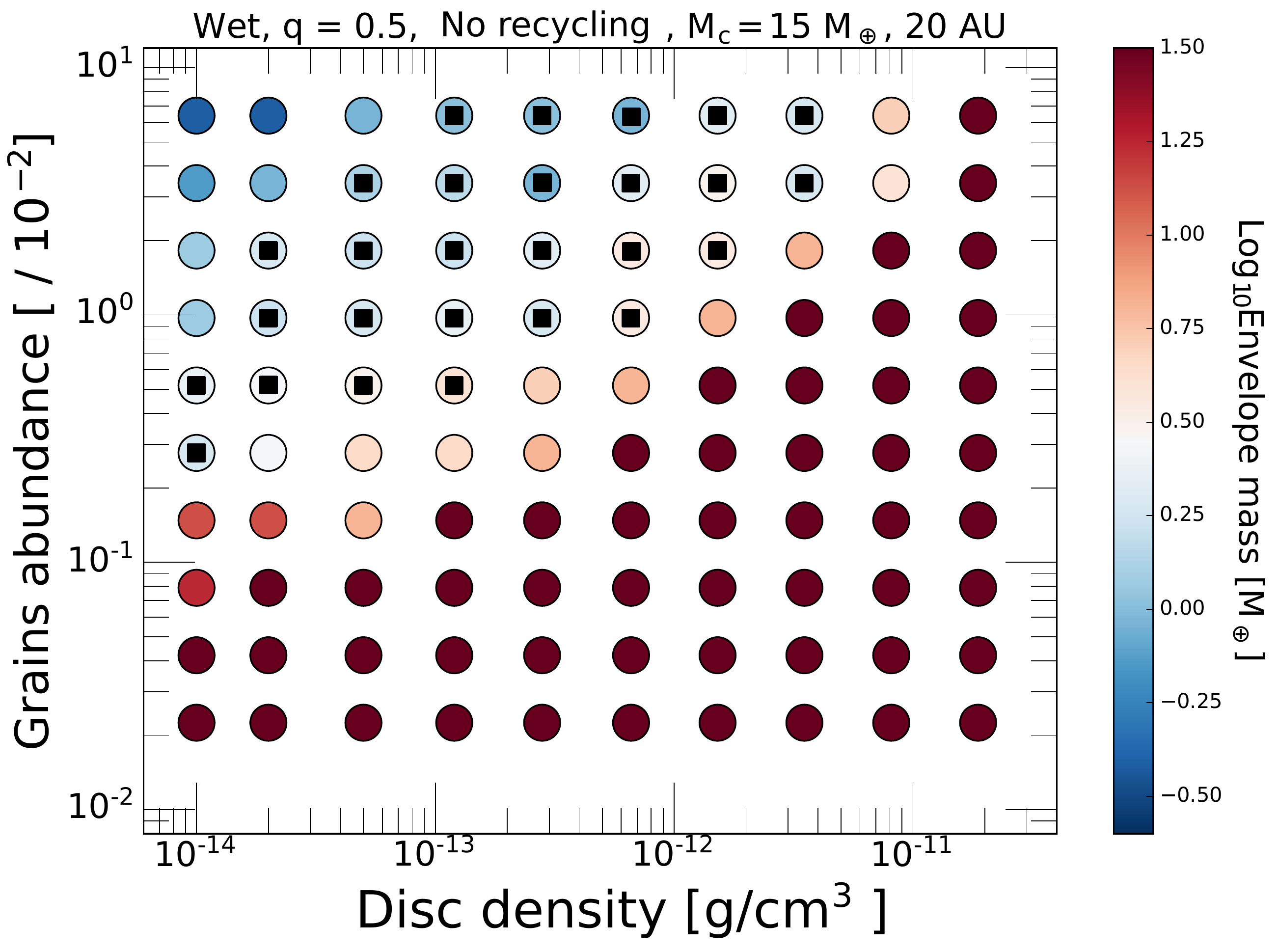}}
\subfloat{\includegraphics[width = 3.5in]{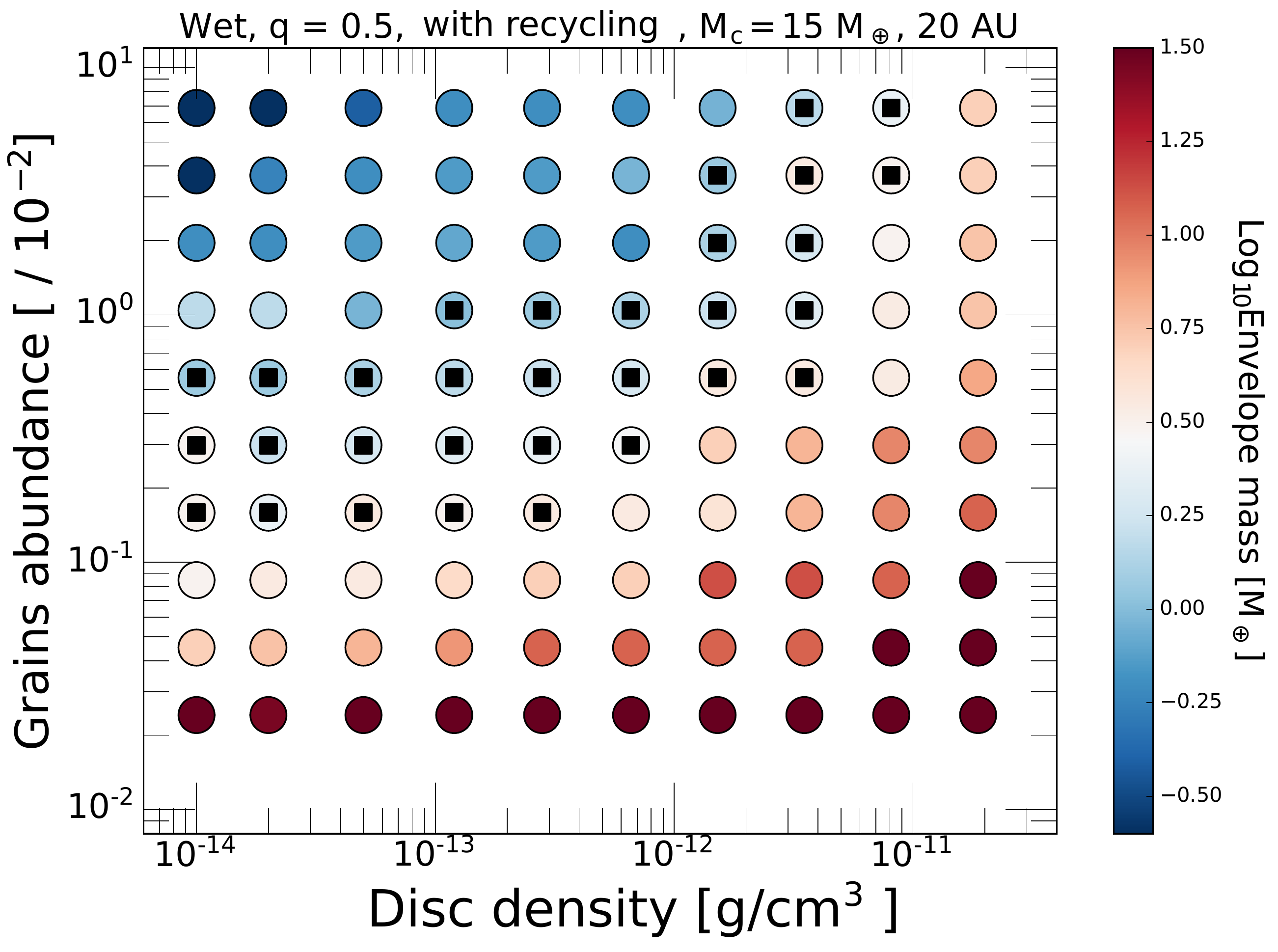}}

\caption{Same as Fig. \ref{fig:results1}, but for a 15 M$_\oplus$ core at 20 AU representing an ice giant. The black markers in this case are the cases with Uranus/Neptune amounts of hydrogen, and where the lower mass cores does not runaway in 10$^6$ yr. }
    \label{fig:resultsUN}
\end{centering}

\end{figure*}

\subsection{Neptune-mass ice giants}
The models we presented here are also relevant for the origin of cold ice giants such as Uranus and Neptune. The mass of solids in Uranus and Neptune is low enough that it can be accommodated without needing a merger. The key issue is whether these planets can attain envelopes with fractional $M_{e}/M_{c} (\sim 0.1)$ prior to the severe depletion of their natal discs.

In Fig. \ref{fig:resultsUN}, we show the results of our simulations but for core mass of 
15 M$_{\oplus}$ at 20 AU where the disc temperature and density are lower than those at 5AU. The simulations are run for the same set of parameters as above, and include the corresponding M$_c$ = 7.5 M$_{\oplus}$ cases. The main inference from this plot is that large parts of parameter space do lead to Uranus-like planets at 20 AU, with wide solution diagonals present in all 4 cases, including those with dry/wet adiabats and with or without entropy advection. Moreover, the results reflect similar trends as those discussed above.

Comparing the two dry cases (i.e.~with and without entropy advection), we find that entropy advection again lowers the final envelope mass and pushes the solution diagonal border line into higher disc densities and lower grain abundances. The wet adiabat also leads to a similar effect, but to a lesser extent. These results are consistent with our \citep{alidib} previous finding that the effects of entropy advection is more significant at 20 than 5 AU. 

Our results are consistent with other Uranus and Neptune formation models. \cite{helled2014} for example ran a series of simulations that include core growth through planetesimal accretion, and gas accretion for a wide range of solid surface surface densities, gas densities, and other free parameters. However, they did not include the effects of wet adiabats or entropy advection.  Our results are best compared to their run \texttt{20UN2} at 20 AU where $\rho_d $ = 8$\times 10^{-12}$ g/cm$^3$, and Z$_{gr}$ is arbitrarily set to be a factor 50 below solar metallicity. (The coexistence of such a massive core with such low grain abundance requires a very efficient mechanism for grain depletion in the envelope.) For these parameters, they obtain a 16.2 M$_\oplus$ core along with a 20 M$_\oplus$ envelope running away in $\sim $0.54 Myr. For the same set of parameters, in Fig.  \ref{fig:resultsUN} (dry case), our protoplanets also reach runaway accretion onto cores with similar masses and on similar timescales. 

\section{Summary and Discussion}
\label{sec:summary}
We investigated the origins of cold sub-Saturns (CSS), planets in the $\sim$ 30-90 M$_\oplus$ mass range which
have been inferred from recent microlensing observations.  These discoveries are in contrast to the anticipation
of a planetary desert in this mass range based on the runaway core accretion scenario of planet formation.  In an 
effort to reconcile theory and observation, we explore the possibility of giant impacts as a mechanism to interrupt the monolithic runaway growth.

By modeling the evolution of protoplanetary envelopes after giant collisions, we show that an impactor in the 0.5-1.5 M$_{\oplus}$ mass range arriving at the escape speed can completely strip a $\sim$ 20-30 M$_{\oplus}$ envelope of an equal mass core completely. 

This process occurs in 2 steps: 1) the impactor's shock wave removes a small fraction of the envelope, and 2) the dissipation of residual kinetic energy increases the core's temperature, leading to super-Eddington luminosities and mass loss. {We emphasize that this thermal mass loss is driven by the core itself, in contrast with the stellar Parker winds considered by \cite{liu2015b}.}

These collisions are more likely to happen during the transitional disc phase due to the secular interaction between multiple planets and their natal discs \citep{nagasawa2005, thommes2008, zheng2017}. { Since eccentric longer-period gas giant planets and companion stars are the most likely perturbers, frequency of giant impacts may be correlated with their occurrence rate which might account for the
suggestive nearly-continuous distribution in the high end (sub-Saturn to super-Jupiter) of the planetary mass function. Further investigations are needed to 
establish both the statistical significance of the observed mass function and the likelihood of giant impacts from relevant population synthesis models.} 

We find that the low-density environment inhibits the re-accreted envelope from reaching runaway accretion during the disc-gas depletion time scale ($\sim 10^5$ yr).  

This growth limit leads to the emergence of the CSS planets.  

From our simulations we infer that:
\begin{enumerate}
    \item It is possible for individual Neptune-mass planets to emerge without runaway at 5 AU after 1 Myr, even with sub-solar grain opacities. These sub-critical Neptune-mass protoplanets are susceptible 
    to envelope-removing impacts during disc dispersal.  
    \item Giant impacts lead to both the core-mass doubling and the ejection of initial gaseous envelope via a combination of shock waves and super-Eddington luminosities.
    \item Although the merged cores are intensely heated, they undergo rapid initial cooling due to the efficient convection in their re-accreted envelope. The timescale of the entire post-collision envelope shedding and core cooling is orders of magnitude less than that of the transitional disc. 
    \item  On the transitional disc-depletion timescale ($\sim 10^5$ yr), the re-accreted envelope's mass is less than that of the merged cores for a wide range of opacities, leading to CSS.
    \item  {Envelope recycling via disc flow slightly reduces the mass of the re-accreted envelopes, but does not significantly modify the results.  }
\end{enumerate}

Based on microlensing statistics, the occurrence rate of CSS exceeds that of cold Jupiters. Although they are less conspicuous in radial velocity surveys, they would be detectable in transitional discs if they are mostly formed through giant impacts. Since giant impacts reset the clock of their evolution, they may have cooling tracks that are inconsistent with the age of their host stars. In addition, giant impacts may lead to byproducts with similar total mass but diverse radius and internal core-envelope structure \citep{liu2015b}.
 
\section*{Acknowledgements}

We thank the anonymous referee for their insightful comments that helped improving this manuscript. We thank Yixian Chen, Eve Lee, Shangfei Liu, Judit Szulágyi, Daniel Thorngren, and Weicheng Zang for useful discussions. AC is supported by an NSERC Discovery Grant and is a member of the Centre de Recherche en Astrophysique du Qu\'ebec (CRAQ) and the Institut de Recherche sur les Exoplan\`etes (iREx). M.A-D is supported through a CAP3 fellowship at NYU Abu Dhabi.

\section*{Data availability}
The data underlying this article (numerical simulations output files) will be shared on reasonable request to the corresponding author.

\bibliographystyle{mnras}

\appendix

\section{Marginal Eddington phase}\label{sec:marginaleddington}

Following a giant collision, we showed in \S \ref{sew} that the luminosity of the core is $\gg L_\mathrm{Edd}$, and we expect super-Eddington winds to thermally strip the core of its envelope entirely, and possibly carve a short lived gap inside the Hill radius. Once the core has cooled enough for this gap to close, so that $L\approx L_\mathrm{Edd}$, radiation pressure might still be relevant, since it will lead to a change in the effective Hill radius. We hence calculate a modified Hill radius accounting for this starting from
\begin{equation}
    \frac{GM_{c}}{R_H^2} - \frac{GM_\ast}{(r-R_H)^2} + \frac{GM_\ast}{r^3}(r-R_H) - 
    \frac{F_\mathrm{rad}}{M_{e}} = 0,
\end{equation}
which can be written as
\begin{equation}
R^3\bigg(\frac{-3M_\ast}{r^3}\bigg) - \frac{F_\mathrm{rad}}{M_{e}G}R^2 + M_{c} = 0.
\end{equation}
For $F_\mathrm{rad}=0$, or small semimajor axis (r in the first term), this reduces to the classic Hill radius. 
For large $F_\mathrm{rad}$, or large semimajor axis, this reduces to
\begin{equation}\label{eq:modHill}
R_H^{Edd} = \bigg(\frac{3M_{c} M_{e} G}{a T_c^4 4 \pi}\bigg)^{1/2} \times R_H^{-1},
\end{equation}
where $T_c$ is the core temperature and $a$ is the radiation constant.

\begin{figure*}
\begin{centering}
        \includegraphics[scale=0.40]{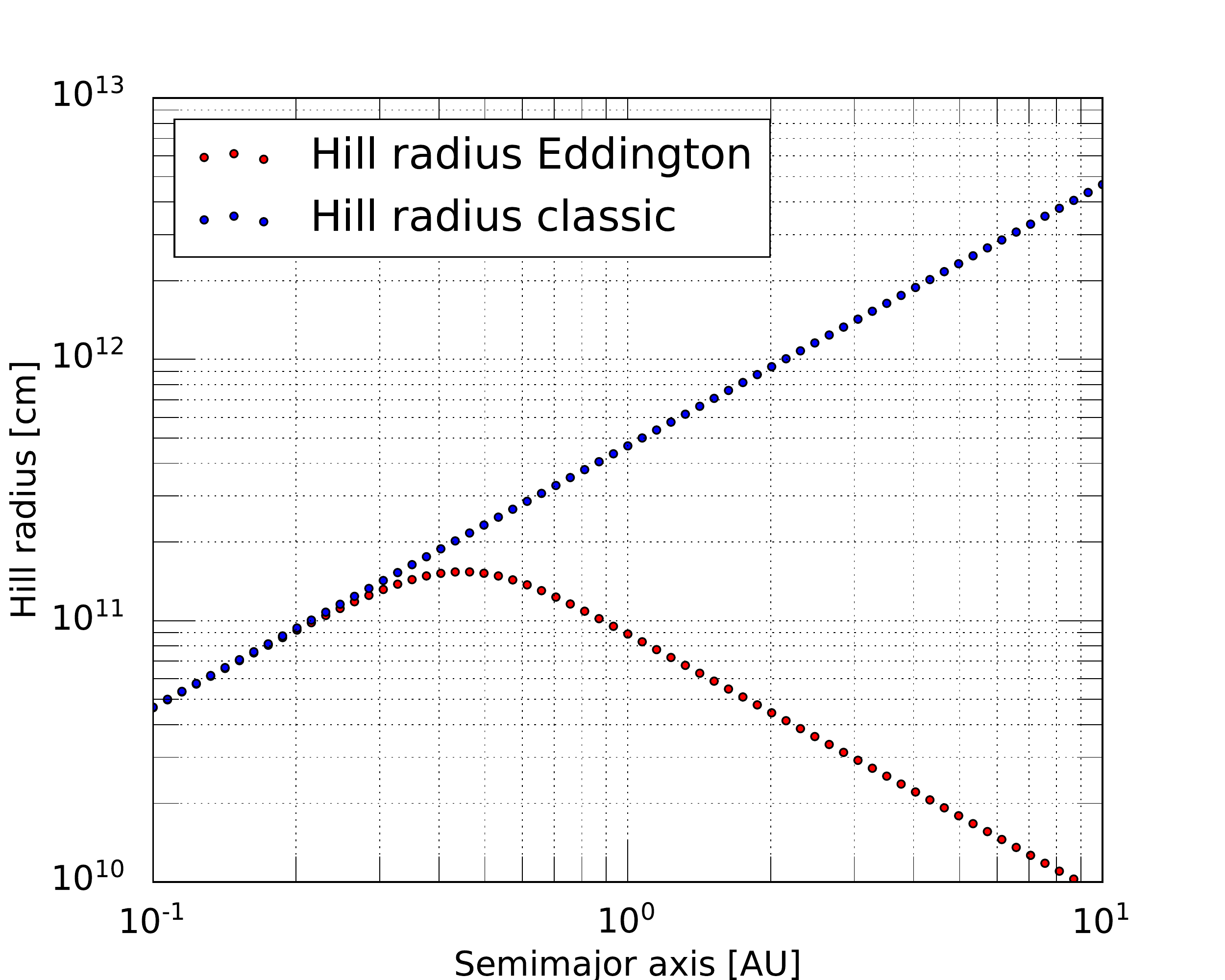}
   \caption{The Hill radius of a hot core with or without including the effects of radiation pressure (eq.~\ref{eq:modHill}) at different orbital radii. We take the core temperature to be T$_\mathrm{core} \sim T_\mathrm{Edd} \sim 4\times 10^4\ {\rm K}$, and the envelope mass to be M$_e \ \sim 0.002 \times M_c$ with $M_c=30\ M_\oplus$.}
    \label{fig:planets2}
    \end{centering}
\end{figure*}

In Fig. \ref{fig:planets2} we show the evolution of the Hill radius for the same nominal parameters (30 M$_{\oplus}$ core) used in sections 2.2, 2.3, and Fig. \ref{fig:planets}, but for T$_c \sim T_\mathrm{Edd} \sim 3\times 10^4\ {\rm K}$. {We assumed M$_e \ \sim 0.02 \times M_c$, which is the envelope mass found when an outer radiative zone has just appeared. This is consistent with the adiabatic envelope masses found by \cite{inamdar} in the same regime. }

Radiation pressure significantly shrinks the Hill radius beyond 1 AU, increasing the cooling time by 2 orders of magnitude at 5 AU. This usually does not affect the re-accretion of the envelope significantly as the system will cool down out of this phase very quickly in the outer disc where the effect is relevant.

\section{The evolution of the envelope's base temperature}

Once the core is cool enough to allow for a radiative zone, the envelope can start to cool down and contract. In our calculations in \S\ref{sec:envelopeaccretion}, we do not include the luminosity from the core, but instead assume that the envelope cooling is dominated by its contraction luminosity. However, in the case that we are considering, the evolution following a giant impact, the core is hot and has significant thermal energy. As it cools, it can act as an internal heat source to the envelope. We investigate this question here and also the time-evolution of the temperature at the base of the envelope.

As the envelope accretes, the temperature at its base can increase or decrease. We can see this with a simple analytic estimate. Consider a massive core (M$_c$, R$_c$) with a low mass envelope (mass $M_e$) such that $M_e\ll M_c$, allowing us to neglect the envelope's self gravity. In this case, from hydrostatic equilibrium the pressure at the base of the envelope is approximately
\begin{equation}
P_b\approx	\frac{GM_{c} M_e}{R_c^4}.
\end{equation}
The rate of change of pressure is
\begin{equation}
    \frac{dP_b}{dt} \approx \frac{GM_c \dot{M}}{R_c^4},
\end{equation}
where $\dot M$ is the accretion rate, giving
\begin{equation}
\label{hse}
\frac{d \ln P_b}{dt} \approx \frac{\dot{M}}{M_e} > 0,
\end{equation}
i.e.~ as the envelope grows, the pressure at the base increases along with its mass.

Now, assuming the envelope is cooling adiabatically, we can also write
\begin{equation}
M_e \Bar{T} \frac{dS}{dt} = - L + L_c,
\end{equation}
where $S$ is the entropy of the envelope, $\Bar{T}$ is the mass-averaged temperature, $L$ the cooling luminosity, and $L_c$ is the luminosity entering the envelope from the core. If we write the entropy as \begin{equation}
S={k_B\over \mu}\ln \left({P\over T^{1/\nabla_\mathrm{ad}}}\right),
\end{equation}
then this gives 
\begin{equation}
{dS\over dt}={k_B\over \mu} \left({d\ln P_b\over dt} - {1\over\nabla_\mathrm{ad}}{d\ln T_b\over dt}\right) = -{L-L_c\over \bar{T} M_e},
\end{equation}
and therefore, using equation (\ref{hse}), 
\begin{equation}\label{eq:dTdt}
    {1\over\nabla_\mathrm{ad}}{d\ln T_b\over dt} = {\dot M\over M_e} -{L-L_c\over k_B\bar{T} M_e/\mu}.
\end{equation}
We see that, whereas the pressure at the base of the envelope always increases as the envelope accretes mass, the temperature at the base could increase or decrease, depending on the relative size of the two terms on the right hand side of equation (\ref{eq:dTdt}). In equilibrium, we would expect $k_B\bar{T}/\mu\sim GM_c/R$, where $R$ is a characteristic radius in the envelope, so the rate of change of temperature depends how the luminosity in the envelope compares to the the accretion luminosity $GM_c\dot M/R$. Neglecting any contribution to the luminosity from the core, we would expect the two terms in equation (\ref{eq:dTdt}) to be comparable, so that the temperature change could be positive or negative.

\begin{figure*}
\begin{centering}
        \includegraphics[scale=1]{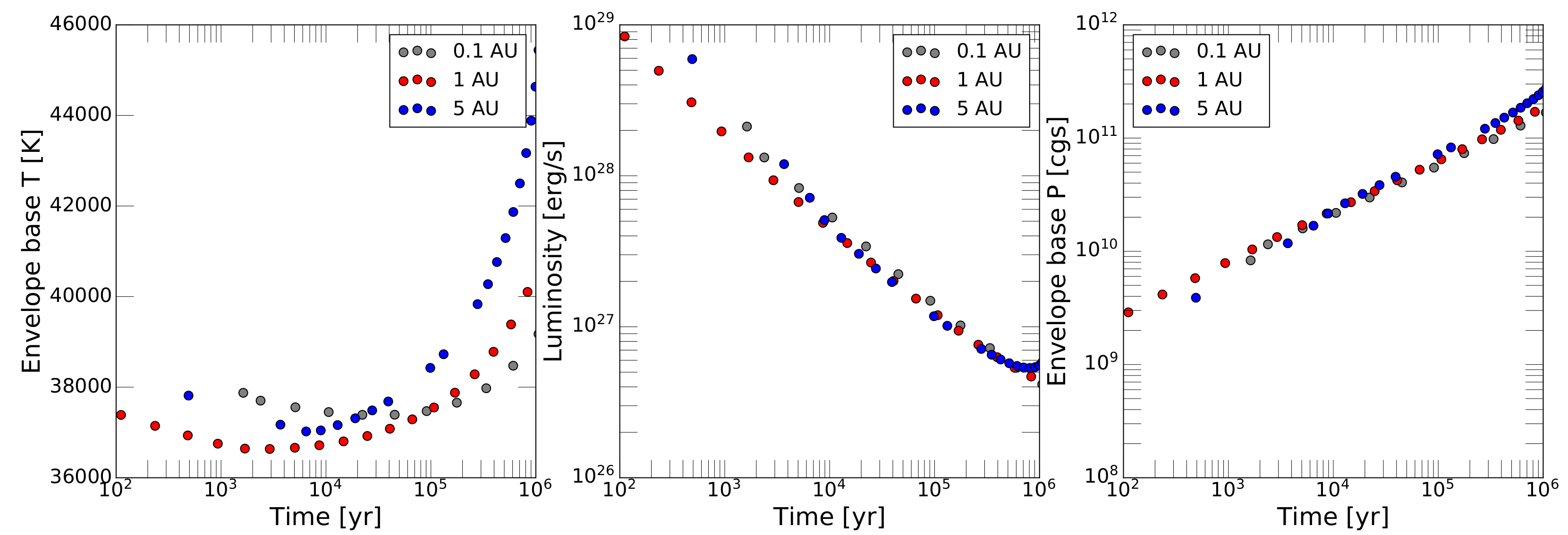}
   \caption{The time evolution of the base temperature (left), luminosity (middle) and base pressure (right) of an accreting envelope around a $30\ M_\oplus$ core at 0.1, 1, and 5 AU. The disc conditions for these three locations are $(\rho_d, T_d)=(5.6\times 10^{-8}\ {\rm g\ cm^{-3}}$, 2000 K), ($6.5\times 10^{-10}\ {\rm g\ cm^{-3}}$, 1010 K), and ($6\times 10^{-11}\ {\rm g\ cm^{-3}}$, 200 K). While the luminosity and pressure change change by two orders of magnitude, the base temperature changes by less than $20$\%.}
    \label{fig:baseT}
    \end{centering}
\end{figure*}

In Fig~\ref{fig:baseT}, using the numerical model outlined in the manuscript, for a 30 M$_\oplus$ core and {assuming ISM grain abundance, we show the envelope's luminosity, base temperature and pressure at 0.1, 1, and 5 AU for the radiatively cooled disc model described in \cite{alidib} (values of disc temperature and density are given in the caption to Fig.~\ref{fig:baseT}).}
We find that in all three cases the base temperature decreases slightly as a function of time before it starts increasing. The base's pressure, on the other hand, always increases monotonically, as discussed in our analytical model above. 

In all cases, if the base temperature is higher than the core's, the envelope will lose energy to the core, slightly accelerating its cooling. If the core is hotter, this adds energy into the envelope, slowing down its cooling. we numerically find that both effects are negligible on timescales larger than $10^3$--$10^4$ yr. This is consistent with a simple estimate assuming the core to be isothermal and have a constant specific heat capacity $C_V=1.129\times 10^7\ {\rm erg\ g^{-1}\ K^{-1}}$ and density $\rho_c=3.3\ {\rm g\ cm^{-3}}$ (as assumed in \S \ref{sec:envloss}). If the core cools according to 
\begin{equation}
    M_c C_V {dT_c\over dt} = -4\pi R_c^2\sigma T_c^4,
\end{equation}
we find $T_c(t)^{-3} = T_i^{-3}+3\sigma t / (R_cC_V\rho_c)$, where $T_i$ is the initial core temperature. Assuming $T_i\gg T_c(t)$ gives  
\begin{equation}
    T_c \approx 2.5\times 10^4\ {\rm K}\ \left({t\over {\rm yr}}\right)^{-1/3} \left({M_c\over 30 M_\oplus}\right)^{1/9},
\end{equation}
with the corresponding luminosity $L_c=4\pi R_c^2 \sigma T_c^4$ or
\begin{equation}
    L_c \approx 7.6\times 10^{27}\ {\rm erg\ s^{-1}}\ \left({t\over 10^4\ {\rm yr}}\right)^{-4/3} \left({M_c\over 30 M_\oplus}\right)^{10/9}.
\end{equation}
Both $T_c$ and $L_c$ are $\propto\rho_c^{2/9}$ at a given time.
Comparing with Fig.~\ref{fig:baseT}, we see that the core luminosity and envelope luminosity are comparable at $t\sim 10^4\ {\rm yrs}$. This assumes that heat is transported efficiently inside the core and so it remains isothermal as it cools. It would be interesting to investigate in more detail the energy transport within the core and across the core-envelope interface for the cases of temperature increasing or decreasing at the core-envelope boundary.

\section{Virial theorem for envelopes}
\label{Virial}

{ In this Appendix, we use the Virial theorem to discuss the expected binding energy and relative size of gravitational and internal energies for an envelope.}

We first give a reminder about the Virial theorem for a star, and then apply it to a planetary envelope. We start with hydrostatic balance ${dP/dr} = -\rho {Gm/r^2}$, where $dm = 4\pi r^2\rho dr = \rho dV$. Multiplying by $r$ and integrating over the volume gives
\begin{equation}
\int 4\pi r^3 dr  {dP\over dr} = -\int {Gm\over r}dm.    
\end{equation}
The right-hand side is the gravitational energy 
\begin{equation}
\Omega = -\int {Gm\over r}dm<0.
\end{equation}
The left-hand side can be integrated by parts as
\begin{equation}\label{eq:byparts}
	\int 4\pi r^3 dr {dP\over dr} = \left[4\pi r^3 P\right] - \int 3P\,4\pi r^2 dr.
\end{equation}
For the case of a star, the surface term vanishes because the integral goes from $r=0$ to $r=R$ where $P=0$. We are left with
\begin{equation}
    \int 3P dV = -\Omega.
\end{equation}
For an adiabatic equation of state with adiabatic index $\gamma$, $P = (\gamma-1)u$, where $u$ is the internal energy density (e.g.~$\gamma=5/3$ gives $P = 2u/3$; $\gamma=4/3$ gives $P = u/3$). Therefore
\begin{equation}\label{eq:virial}
\int 3P dV = \int 3(\gamma-1)u\, dV = 3(\gamma-1)U = -\Omega,
\end{equation}
where we write the total internal energy as $U$. 

Equation (\ref{eq:virial}) is the usual Virial theorem for stars (neglecting rotation and time-dependence). For $\gamma=5/3$, we therefore have $-\Omega/U = 3(\gamma-1) = 2$, which is the familiar result that the internal energy is one half of the gravitational energy. The total energy is
$$E_{\rm tot} = \Omega + U = \Omega \left({\gamma-4/3\over \gamma-1}\right).$$
For $\gamma=5/3$, $E_{\rm tot} = \Omega/2 <0$ as expected. For $\gamma=4/3$, the magnitude of the internal energy and gravitational energy become equal, and $E_{\rm tot}=0$. A star with $\gamma<4/3$ would be unbound. 

Now consider an adiabatic envelope on a core with mass $M_c$, radius $R_c$ and volume $V_c=4\pi R_c^3/3$. The argument is the same as before, but we now need to keep the surface term in equation (\ref{eq:byparts}). This gives
\begin{equation}\label{eq:appC1}
    4\pi R_c^3 P_c + 3(\gamma-1)U = -\Omega,
\end{equation}
where the integrals in $\Omega$ and $U$ are evaluated over the volume of the envelope, and $P_c$ is the pressure at the base of the envelope (at $r=R_c$).

{Equation (\ref{eq:appC1}) can be written as:
\begin{equation}
\frac{U}{\Omega} = -\frac{1}{3(\gamma-1)} - \frac{V_c P_c}{\Omega(\gamma-1) }
\end{equation}
allowing us to evaluate the internal to gravitational energy ratio analytically, and compare our analysis above to the numerically calculated ratio. For a 30 M$_\oplus$ core with equal-mass envelope at 5 AU, the numerical simulation U/$\Omega$ ratio from Table \ref{tablenummimm} is -0.70 and -0.95 respectively for $\gamma$ = 1.4 and 1.25. Inserting the analytically calculated $\Omega$ ($\equiv$ $E_G^a$) in Table \ref{tablenummimm} and using $P_c$ from the numerical simulations, we get U/$\Omega$ ratios of -0.78 and -1.09 respectively for $\gamma$ = 1.4 and 1.25. Our analytical analysis in hence consistent with the results of the numerical simulations.}

We moreover see that the total energy now has an extra term in it,
\begin{equation}\label{eq:Etot}
E_{\rm tot} = \Omega \left({\gamma-4/3\over \gamma-1}\right) -{P_cV_c\over \gamma-1}.	
\end{equation}
This means that we can have envelopes with $\gamma < 4/3$ because even though the first term is then positive, there is an additional negative term coming from the core that can keep the envelope bound. The minimum value of $\gamma$ for a bound envelope is
\begin{equation}\label{eq:gmin}
	\gamma_{\rm min} = {4\over 3} -{P_cV_c\over (-\Omega)}.
\end{equation}
In the limit of small envelope mass $\Delta M\ll M_c$ and a thin envelope, we can estimate 
\begin{equation}
    P_c\approx {GM_c\Delta M\over R_c^2}{1\over 4\pi R_c^2}
\end{equation} and $-\Omega \approx (GM_c/R_c)\Delta M$. This gives $P_cV_c/(-\Omega) = 1/3$, so that $\gamma_{\rm min}=1$ in this limit, corresponding to an isothermal atmosphere.



\bsp	
\label{lastpage}

\end{document}